\documentclass[conference]{IEEEtran}
\usepackage[most]{tcolorbox}
\usepackage[switch,mathlines]{lineno}

\usepackage{graphicx}
\usepackage{balance}  %
\usepackage{float}
\usepackage{multirow}
\usepackage{caption}
\usepackage{enumitem}

\usepackage[linesnumbered,ruled,vlined,noend]{algorithm2e}
\SetKwInput{KwInput}{Input}                %
\SetKwInput{KwOutput}{Output}              %
\newcommand{\algoFontSize}{\small} %

\usepackage{mdframed}
\newmdenv[
  topline=false,
  bottomline=false,
  skipabove=\topsep,
  skipbelow=\topsep,
  leftmargin=-10pt,
  rightmargin=-10pt,
  innertopmargin=0pt,
  innerbottommargin=0pt
]{siderules}

\usepackage{wrapfig}

\usepackage{hyperref}

\usepackage{amsmath}

\usepackage{xcolor}
\usepackage{bm}

\usepackage{amssymb}
\def\ojoin{\setbox0=\hbox{$\bowtie$}%
  \rule[-.02ex]{.25em}{.4pt}\llap{\rule[\ht0]{.25em}{.4pt}}}
\def\leftouterjoin{\mathbin{\ojoin\mkern-5.8mu\bowtie}}
\def\rightouterjoin{\mathbin{\bowtie\mkern-5.8mu\ojoin}}
\def\fullouterjoin{\mathbin{\ojoin\mkern-5.8mu\bowtie\mkern-5.8mu\ojoin}}

\usepackage{tikz}
\usetikzlibrary{shapes,arrows}

\newcommand{\thmFont}[0]{\it}
\newcommand{\defFont}[0]{\it}

\newcommand{\textRev}[1] {#1} %

\newcommand{\jedi}{{\small\textsf{InFine~}}}

\newcommand{\jfd}{{\small\textsf{Inferred FD discovery~}}}

\newcommand{\leftInst}[0]{\ensuremath{L}\xspace}
\newcommand{\rightInst}[0]{\ensuremath{R}\xspace}
\newcommand{\leftSch}[0]{\ensuremath{\mathbf{S}}\xspace}
\newcommand{\rightSch}[0]{\ensuremath{\mathbf{T}}\xspace}
\newcommand{\leftJoinAtt}[0]{\ensuremath{X}\xspace}
\newcommand{\rightJoinAtt}[0]{\ensuremath{Y}\xspace}

\newcommand{\view}[0]{\ensuremath{V}\xspace}
\newcommand{\leftView}[0]{\ensuremath{V_\leftInst}\xspace}
\newcommand{\rightView}[0]{\ensuremath{V_\rightInst}\xspace}
\newcommand{\viewSpec}[0]{\ensuremath{{\view_{\relationSet}}}\xspace}
\newcommand{\viewSpecArg}[1]{\ensuremath{{\view_{#1}}}\xspace}
\newcommand{\fdSet}[0]{\ensuremath{\mathcal{D}}\xspace}
\newcommand{\relationSet}[0]{\ensuremath{\mathbf{R}}\xspace}
\newcommand{\relationsFDsSets}[0]{\ensuremath{\mathcal{E}_\fdSet}\xspace}
\newcommand{\projectedAttsSet}[0]{\ensuremath{\mathcal{A}_\view}\xspace}
\newcommand{\provenanceSet}[0]{\ensuremath{\mathcal{P}}\xspace}

\newcommand{\fun}[1]{\ensuremath{\mathrm{#1}}\xspace}

\newcommand{\atts}[0]{\ensuremath{atts}}
\newcommand{\fds}[0]{\ensuremath{fds}}
\newcommand{\projectedAttributes}[0]{\ensuremath{proj}}

\newcommand{\add}{{\bf add}\xspace}
\newcommand{\get}{{\bf get}\xspace}
\newcommand{\compute}{{\bf compute}\xspace}
\newcommand{\prune}{{\bf prune}\xspace}

\newcommand{\provFDs}{\texttt{provFDs}\xspace}
\newcommand{\compUpstaged}{\texttt{joinUpFDs}\xspace}
\newcommand{\subInfer}{\texttt{infer}\xspace}
\newcommand{\subRefine}{\texttt{refine}\xspace}

\newtheorem{definition}{Definition}

\newtheorem{theorem}{Theorem}
\newtheorem{lemma}{{Lemma}}

\newtheorem{example}{Example}

\newcommand {\approxi}[1]{\rightharpoondown^{#1}}

\begin{document}

\title{Provenance-aware Discovery of Functional Dependencies on Integrated Views}
\author{
\IEEEauthorblockN{Ugo Comignani}
\IEEEauthorblockA{Tyrex team, Grenoble INP,\\
INRIA, France\\
ugo.comignani@inria.fr}
\and
\IEEEauthorblockN{Laure Berti-Equille}
\IEEEauthorblockA{IRD, ESPACE-DEV\\
Montpellier, France \\
laure.berti@ird.fr}
\and
\IEEEauthorblockN{Noël Novelli}
\IEEEauthorblockA{Aix-Marseille Univ.\\
LIS CNRS, France \\
noel.novelli@lis-lab.fr}
\and
\IEEEauthorblockN{Angela Bonifati}
\IEEEauthorblockA{ Lyon 1 University\\
Lyon, France \\
angela.bonifati@univ-lyon1.fr }
}

\maketitle

\begin{abstract}
The automatic discovery of functional dependencies (FDs) has been widely studied as one of the hardest problems in data profiling. Existing approaches have focused on making the FD computation efficient while inspecting single relations at a time. In this paper, for the first time we address the problem of inferring FDs for multiple relations as they occur in integrated views by solely using the 
functional dependencies of the base relations of the view itself. 
To this purpose, we leverage logical inference and selective  mining %
and show that we can discover most of the exact FDs from the base relations and avoid the full computation of the FDs for the integrated view itself, while at the same time preserving the lineage of FDs of base relations. We propose algorithms to speedup the inferred FD discovery process and mine FDs on-the-fly only from necessary data partitions. We present \jedi ({\small{\it INferred FunctIoNal dEpendency}}), an end-to-end solution to discover inferred FDs on integrated views by leveraging provenance information of base relations. %
Our experiments on a range of real-world and synthetic datasets demonstrate the benefits of our method over existing FD discovery methods that need to rerun the discovery process on the view from scratch and cannot exploit lineage information on the FDs. 
 We show that \jedi outperforms traditional methods necessitating the full integrated view computation by one to two order of magnitude in terms of runtime. It is also the most memory efficient method while preserving FD provenance information using mainly inference from base table with negligible execution time. 

\end{abstract}

\section{Introduction}

The automatic discovery of all functional dependencies (FDs) holding in a single relation is amongst the hardest problems in 
data profiling \cite{2018Abedjan,PapenbrockEMNRZ15}. 
Typically, an FD $X \rightarrow Y$ with attribute sets $X$ and $Y$ in a given table allows to enforce that the combination of values in the set $X$ uniquely determines the values of every attribute in the set $Y$. Functional dependencies are key ingredients in database design, table decomposition, database normalization, and for several other data management tasks, such as data cleaning \cite{Thirumuruganathan17} 
and query optimization \cite{IlyasMHBA04, Paulley}.  
\textRev{Due to the high complexity of FD discovery \cite{Liu2012} (exponential in the number of attributes and quadratic in the number of records of a relation), a wealth of algorithms have been proposed by relying on aggressive pruning and sophisticated validation. Many of these algorithms \cite{HKPT98,HKPT99,WGR01,NoCi01icdt,HyFD} have the same theoretical complexity and are able to compute all minimal FDs in less runtime.}  

Despite the vast literature on the topic, all existing algorithms focus on the FD discovery problem for one relation at a time \textRev{and do not explore reuse of computation when dealing with multiple relations}. 
However, relations are oftentimes involved in view computation operations, \textRev{among which  
the notable class of SPJ (Select-Project-Join) view operations}. 
In this case, the existing algorithms addressing FD discovery would have to entirely recompute the set of FDs for the obtained integrated view without being able to reuse any of the previous computations on the individual base tables of the view. 
In this paper, for the first time we tackle the problem of inferring FDs on integrated views (also called \jfd problem) by solely relying on the FDs available from the base tables of the views. To do so, we employ the well-known concept of why-provenance \cite{BunemanKT01}, which allows us to 
identify the lineage of the FDs on the integrated view and throughout the view computation operations.

 \begin{figure*}[t]
	\centering
	\includegraphics[width=1.05\linewidth]{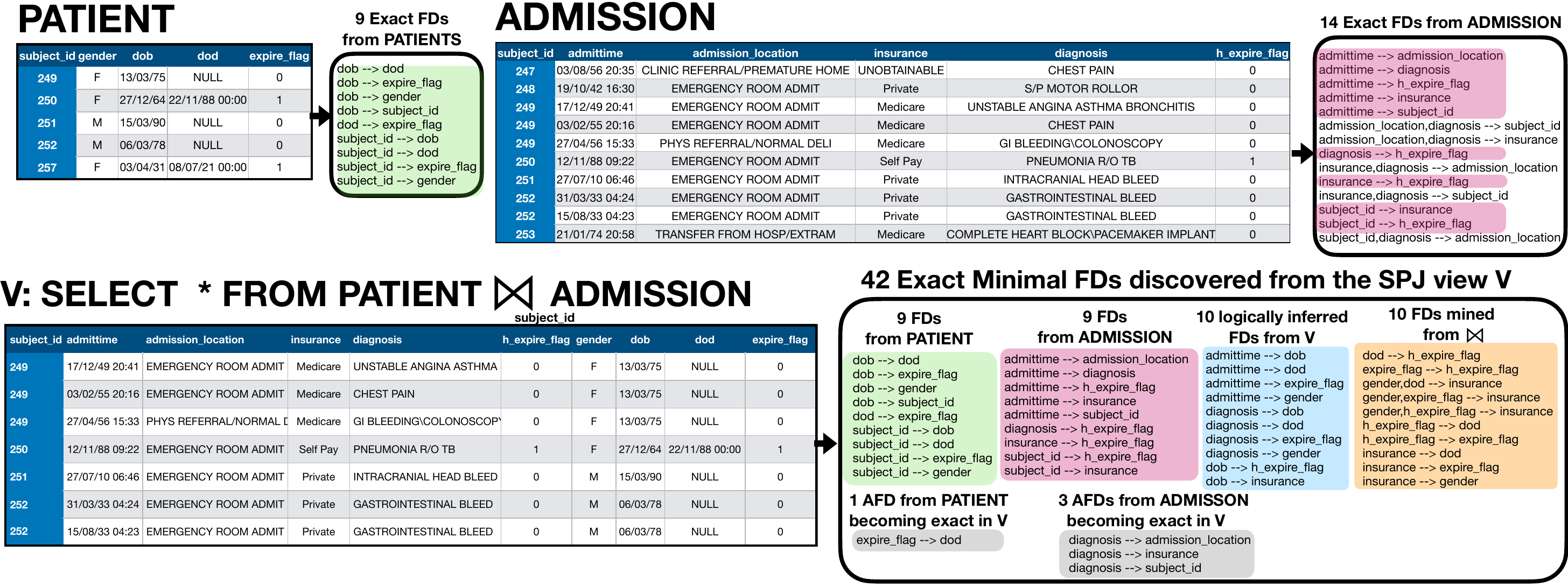}
	\caption{Excerpt of the MIMIC-III clinical database with FDs on base tables and integrated view (best viewed in color).}\label{fig:example}
\end{figure*} 

\textRev{
In our work, we address the \jfd problem, which consists of considering a SPJ view on top of a set of base tables and computing the FDs holding on the view by reusing as much as possible the discovered FDs on the base tables, and, in turn, reduce the SPJ view computation with only the needed attributes. 

Explicitly relating the FDs of the base table to the FDs of the view has interest on its own as it allows for instance to understand whether FDs on the base tables are persistent on the view. It can also be beneficial whenever the user needs help while debugging the FDs on the base tables by choosing the most relevant ones if they also apply to the view. Since SPJ views might be obtained as results of data integration and ETL scenarios \cite{GolshanHMT17}, our method allows to understand how constraints (namely the FDs) are actually affected by the integration process. New FDs that hold on the view but not on the base tables may help the user better understand and explain the result of the data integration process. 

Our method allows us to obtain time savings in terms of FD discovery from integrated views. By leveraging logical inference and provenance triples, we can avoid recomputing the FDs holding on the view from scratch and only focus on the new FDs that are not holding on the base tables.}

To address the \jfd problem, we use a multi-step pipeline encompassing the discovery of different types of FDs %
 and leveraging logical inference.  
 To this end, we propose {\textsf{InFine}}, an efficient solution for automatically discovering multi-relation FDs starting from the FDs of the base tables. 
Our main contributions are as follows:
\begin{itemize}[noitemsep, nolistsep, leftmargin=*,topsep=0pt]
    \item We propose a provenance-based mechanism 
    capable of generating and exploiting provenance triples in a view specification. In particular, for each FD, we capture the type of the FD and a subquery of the view in which the FD holds. 
    \item We design five algorithms to compute the provenance of FDs (from base FDs, inference, or join operations) from the results of the FD discovery on the base tables; these algorithms seamlessly address the case of SPJ views, a significant and representative query fragment. 
    \item We propose {\textsf{InFine}}, %
     a full-fledged system implementing our algorithms. \jedi is available online\footnote{\url{https://github.com/ucomignani/InFine}} with code, scripts, and datasets for the reproducibility of our experiments;
    \item We gauge the effectiveness of our system through an extensive experimental evaluation. We compare \jedi against state-of-the-art FD discovery methods over a rich number of SPJ views on real-world and synthetic datasets. We find that \jedi outperforms the competing methods by one order up to two orders of magnitude in terms of execution time for discovering exact FDs while preserving the smallest memory consumption on average.
\end{itemize}

\noindent {\bf Outline.} Section~\ref{sec:example} presents an illustrative example. Section~\ref{sec:preliminaries} presents the necessary background and notations. \textRev{In Section~\ref{sec:contributions}, we formalize the \textsf{Inferred FD discovery~}problem and provide an overview of {\textsf{InFine}}. We also present our main contributions and the algorithms at the core of {\textsf{InFine}}.} We describe our performance experiments evaluating the efficiency and accuracy of {\textsf{InFine}} in  Section~\ref{sec:experiments}. Finally, we discuss related work in Section~\ref{sec:relatedwork} and conclude in Section~\ref{sec:conclusion}.

\section{Motivating Example}
\label{sec:example}

Let us consider a real-world
clinical database \texttt{ MIMIC-III}\footnote{\url{https://physionet.org/content/mimiciii/1.4/}} \cite{mimiciii}, out of which we extract the  {\small\texttt{PATIENT}} table containing information about patients: their identifier {\small\texttt{(subject\_id)}}, \texttt{gender}, date of birth {\small\texttt{(dob)}}, date of death {\small\texttt{(dod)}}, and a boolean \mbox{{\small\texttt{expire\_flag}}} indicating whether the patient passed away,  and the table  {\small\texttt{ADMISSION}} containing administrative and clinical  information about patients such as the hospital admission time {\small\texttt{(admittime)}}, the admission location {\small\texttt{(admission\_location)}}, the {\small\texttt{insurance}}, the {\small\texttt{diagnosis}}, and a boolean {\small\texttt{h\_expire\_flag}}  indicating whether the patient died at the hospital. %
Let us consider $V$, the SPJ view  (illustrated in Fig. 1) that computes a join between the two above tables: {\small\texttt{SELECT * FROM PATIENT, ADMISSION}} {\small\texttt{WHERE PATIENT.subject\_id=ADMISSION.subject\_id}}. We are interested in discovering the set of FDs that hold over the SPJ view  by reusing as much as possible the FDs of the base tables {\small\texttt{PATIENT}} and {\small\texttt{ADMISSION}}. \textRev{For ease of exposition,  we focus on FDs with RHS limited to one single attribute (as it is the case in a canonical cover of FDs). The LHS might consist of multiple attributes as those mined from the join in Fig. 1. A total of 42 minimal and canonical FDs is considered in this running example thus making the example far from being trivial. These FDs can be obtained by FD discovery methods running on the view. However, their lineage is also important as they carry the information of whether they are valid in the base tables or they solely hold on  the view. %
} 
Precisely, we use color coding to encode the FD provenance in the figure.  We can observe that the 9 exact FDs from  the base table {\small\texttt{PATIENT}} (highlighted in green), as well as 9 (out of the 14) exact FDs from  the base table {\small\texttt{ADMISSION}} (highlighted in pink) are preserved in the result of the integrated view. \textRev{Imagine that a data steward would like to investigate why a few valid FDs in the base tables are no longer valid in the integrated view result or, conversely, why some FDs that are not valid in the base tables become valid in the view. Better informed, s/he may change and adapt the constraint (FD) enforcement strategies when curating the data using the FDs.} 
\textRev{Moreover, 10 FDs (highlighted in blue) from the view result can be obtained by logical inference over the sets of exact FDs discovered from each base table: for instance,
${\small\texttt{diagnosis}}\to {\small\texttt{dod}}$ is obtained from ${\small\texttt{insurance,diagnosis}}\to  {\small\texttt{subject\_id}}$  and ${\small\texttt{subject\_id}}\to  {\small\texttt{insurance}}$  in {\small\texttt{ADMISSION}} and  ${\small\texttt{subject\_id}} \to  {\small\texttt{dod}}$ in {\small\texttt{PATIENT}}. As a side note, we are not addressing here the problem of meaningfulness of the FDs, which is orthogonal and of independent interest. %
 Indeed,  a valid FD may not be semantically meaningful and judging whether a valid FD is relevant typically needs human intervention.} 

When the two tables are joined, patient \#257 is removed due to the absence of the corresponding  ${\small\texttt{subject\_id}}$ value in the other table and  approximate FDs (AFD) (such as ${\small\texttt{expire\_flag}} \approxi{1} {\small\texttt{dod}}$ in {\small\texttt{PATIENT}}) become exact in the SPJ view. These FDs are highlighted in grey in the figure. 
 
Finally, only 10 exact FDs (highlighted in orange) that hold over the view result have to be discovered from scratch from the view. However, if we partially join the two tables, only with the following combinations of tuples: [(\#249,\#252) or (\#249,\#251)] and [(\#250,\#251) or (\#250, \#252)], we can obtain the remaining 10 join FDs without having to compute the entire view beforehand. 

This example show that with existing FD discovery approaches, the FDs would have to be computed on both the base tables and the integrated view result \textRev{to preserve the FD provenance information}. Furthermore, to identify the provenance of the FDs, a comparison among the two FD sets would have to be performed. We will show the overhead of this process in our experimental study in Section V.

Hence, a better understanding of the mechanisms underlying the provenance of FDs from integrated views is highly needed. Based on these observations, in our work,  we address the following key questions: How can we preserve the provenance of FDs discovered from integrated views? Instead of executing FD discovery over each base table and a SPJ view result independently, can we infer most of the FDs on the view as well as reuse the FDs from the base tables and achieve a non negligible speedup? We answer these key questions in the rest of the paper and propose an efficient solution for discovering FDs from integrated views.

\section{Preliminaries}
\label{sec:preliminaries}

Next, we recall the necessary definitions of FDs and join operators with their application to our problem.
\begin{definition}[Functional dependency satisfaction]\defFont
  Let $I$ be an instance over a relation schema $R$, 
  and $\bm{X}$, $\bm{Y}$ be two sets of attributes from $R$. $I$ satisfies a functional dependency $d: \bm{X} \to \bm{Y}$, denoted by  $I \models d$ if and only if:
  \begin{equation}
  \forall t_1,t_2\in I, t_1[\bm{X}] = t_2[\bm{X}] \Rightarrow t_1[\bm{Y}] = t_2[\bm{Y}].      
  \end{equation}
 \end{definition}
 \textRev{Note that the above definition holds regardless of the selected null semantics. This implies that our approach is not depending on the underlying semantics of null values, in line with other approaches in the literature \cite{halloween2018,DurschSWFFSBHJP19,Kruse18}.
 
 Following the convention, we use uppercase letters for attribute sets and lowercase letters for single attributes. Moreover, in the rest of the paper we use canonical FDs, i.e., minimal FDs with only one attribute in their right-hand part.%
  We recall that for any set of FDs, it can be computed a logically equivalent set of canonical FDs, thus this is done without loss of generality.
 }
We define the SPJ view specification and the set of projected attributes over this view as follows:
\begin{definition}[SPJ view specification]\defFont
Let $\relationSet = \{R_1;\dots;R_n\}$ be a set of relational instances.
We define a view specification $\viewSpec$ as a relational algebra formula over relations in $\mathbf{R}$ and limited to the following set of operators: $\{\pi;\sigma;\bowtie;\fullouterjoin;\leftouterjoin;\rightouterjoin;\ltimes;\rtimes\}$.
\end{definition}
\begin{definition}[Projected attributes set]\defFont
Let $\viewSpecArg{}$, $\viewSpecArg{1}$, and $\viewSpecArg{2}$ be view specifications.
Let $R$ be a relational instance.
Let $X$ be a set of attributes.
Let $\rho$ be a set of constraints.
Let $\atts(S)$ denotes the set of attributes over a relational instance $S$.
Let $\diamond$ be a join operator in $\{\bowtie;\fullouterjoin;\leftouterjoin;\rightouterjoin\}$.
Then the set of projected attributes, denoted by  $\projectedAttributes()$ is defined as follows:
\begin{align*}
 &\projectedAttributes(R) = \atts(R) &&  \projectedAttributes(\viewSpecArg{1} \ltimes \viewSpecArg{2}) = \projectedAttributes(\viewSpecArg{1})\\
 &\projectedAttributes(\pi_X(\viewSpecArg{})) = X && \projectedAttributes(\viewSpecArg{1} \rtimes \viewSpecArg{2}) = \projectedAttributes(\viewSpecArg{2}) \\ 
 &\projectedAttributes(\sigma_\rho(\viewSpecArg{})) = \projectedAttributes(\viewSpecArg{}) && \projectedAttributes(\viewSpecArg{1} \diamond \viewSpecArg{2}) = \projectedAttributes(\viewSpecArg{1}) \cup \projectedAttributes(\viewSpecArg{2})
\end{align*}

\end{definition}

\textRev{Finally, we define the types of FDs and the provenance triples derived in our framework and used to maintain the provenance information of every discovered FD.
The base FDs are defined as follows:
\begin{definition}[Base FD]
Let $\relationSet = \{R_1;\dots;R_n\}$ be a set of relational instances.
Let $\viewSpec$ be a view specification over relations in $\mathbf{R}$.
Let $\fds(R_i)$ denotes the set of minimal FDs over the relation $R_i \in \relationSet$.
Let $\fds(\viewSpec)$ denotes the set of minimal FDs over the view specified by $\viewSpec$.
An FD $d$ is a base FD if $d \in \fds(\viewSpec)$ and $\exists R_i\in\relationSet \text{ s.t. } d\in\fds(R_i)$.
\end{definition}
We now define the notion of upstaged FD, which can occurs either in the case of a selection or of a join operation:
\begin{definition}[Upstaged FD]
Let $R_1,R_2$ be two relational instances.
Let $\fds(R_i)$ denotes the set of minimal FDs over a relational instance $R_i$.
An FD $d$ is an upstaged FD for $R_1$ if $d$ is defined over attributes from $R_1$ and:
\begin{itemize}
\item in the case of a selection $\sigma_\rho(R_1)$: $d \in \fds(\sigma_\rho(R_1))$ and $d \not\in \fds(R_1)$
\item in the case of a join $R_1 \diamond R_2$: $d \in \fds(R_1 \diamond R_2)$ and $d \not\in \fds(R_1)$
\end{itemize}
\end{definition}
In the following, we distinguish the upstaged FDs by considering if they come from a selection operation or a join operation. 

We also define two types of FDs that arise specifically after a join operation is performed:
\begin{definition}[Inferred FD]
Let $R_1,R_2$ be two relational instances.
Let $\fds(R_i)$ denotes the set of minimal FDs over a relational instance $R_i$.
Let $\atts(R_i)$ denotes the set of attributes of $R_i$.
Let $d: X \to y$ be a FD such that $d\in\fds(R \diamond R^\prime)$.
Then $d$ is an inferred FD if
$$ (X\cup\{y\}) \cap \atts(R_1) \neq \emptyset \wedge (X\cup\{y\}) \cap \atts(R_2) \neq \emptyset$$
and either:
\begin{itemize}
    \item $d$ can be inferred through the use of Armstrong's axioms;
    \item $\exists d^\prime: X^\prime \to y$ such that $X\subset X^\prime$ and $d^\prime$ can be inferred through Armstrong's axioms.
\end{itemize}
\end{definition}
\begin{definition}[Join FD]
Let $d: X \to y$ be a FD such that $d\in\fds(R \diamond R^\prime)$.
Then $d$ is a join FD if
$$ (X\cup\{y\}) \cap \atts(R_1) \neq \emptyset \wedge (X\cup\{y\}) \cap \atts(R_2) \neq \emptyset$$
and $d$ did not belong to the set of inferred FDs.
\end{definition}
From these types of FDs, we define the provenance triples as follows:
}
\begin{definition}[FD Provenance Triple]\defFont
Let $\relationSet = \{R_1;\dots;R_n\}$ be a set of relational instances.
Let $\viewSpec$ be a view specification over relations in $\mathbf{R}$.
A provenance triple $(d,t,s)$ for an FD $d$ over the view specified by $V_{\mathbf{R}}$ is a triple composed of:
\begin{itemize}
    \item the FD $d$ whose provenance is described;
    \item the type $t$ of $d$, taking one of the following values: $``base"$, $``upstaged\ selection"$, $``upstaged\ left"$, $``upstaged\ right"$, $``inferred"$ or $``join FD"$;
    \item \textRev{the first sub-query $s$ over the view specification $V_{\mathbf{R}}$ in which $d$ holds during the view computation.}
\end{itemize}
\end{definition}
\textRev{We illustrate the provenance triples in the following example:
\begin{example}
Following the running example of Figure~\ref{fig:example}, the provenance triple for FDs $\text{\small\texttt{subject\_id}} \to \text{\small\texttt{dob}}$; $\text{\small\texttt{expire\_flag}} \to \text{\small\texttt{dod}}$ and $\text{\small\texttt{gender, h\_expire\_flag}} \to \text{\small\texttt{insurance}}$
will be the following:
\begin{align*}
 (&\text{\small\texttt{subject\_id}} \to \text{\small\texttt{dob}},"base", \text{\small\texttt{ADMISSION}})\\
 (&\text{\small\texttt{expire\_flag}} \to \text{\small\texttt{dod}}, "upstaged\ left",\\
  &\qquad\text{\small\texttt{PATIENT}} \bowtie_{subject\_id = subject\_id}, \text{\small\texttt{ADMISSION}})\\
 (&\text{\small\texttt{gender, h\_expire\_flag}} \to \text{\small\texttt{insurance}},"join\ FD",\\
  &\qquad\text{\small\texttt{PATIENT}} \bowtie_{subject\_id = subject\_id} \text{\small\texttt{ADMISSION}}).
\end{align*}

\end{example}
}%

 \begin{figure}[t]
	\centering
	\includegraphics[width=.9\linewidth]{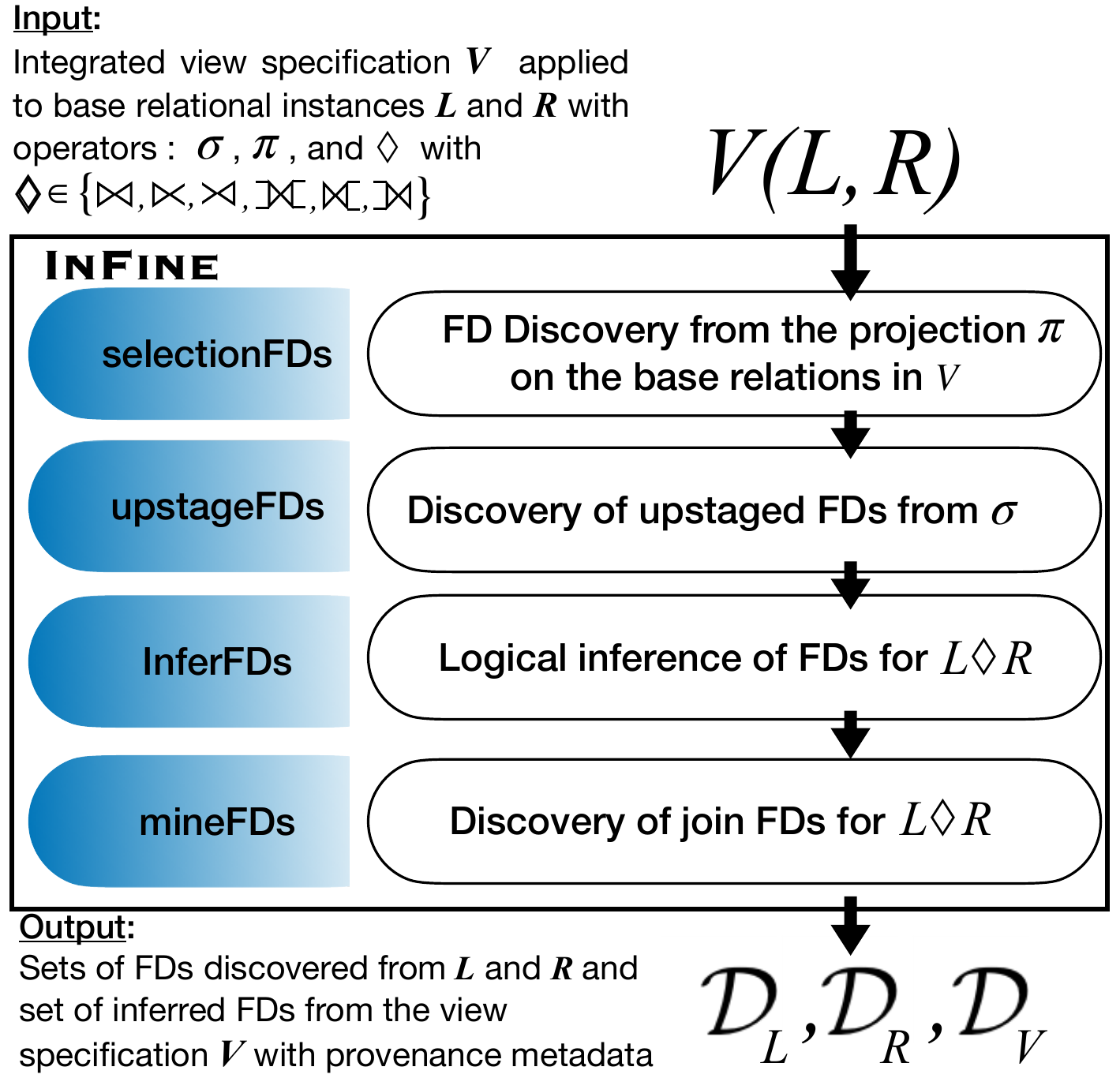}
	\caption{Workflow of \jedi for discovering inferred FDs from  a view specification $V(L,R)$%
	}\label{fig:workflow}
\end{figure} 

\section{Mining FDs provenance triples over Views}
\label{sec:contributions}
In this section, we describe the workflow of \jedi to compute FDs and their provenance information from a view specification and the base relations used in the view specification.
In this workflow, after discovering FDs from the base tables, we mine the FDs appearing during the view computation by relying as much as possible on inference methods as well as on efficient methods to discover the remaining FDs that cannot be inferred.

{\bf Problem statement}.\label{prob:generalJfds}
\textRev{{\it Let $\relationSet$ be a set of base tables;
\viewSpec be a view specification over \relationSet.
The \jfd problem consists of efficiently inferring functional dependencies over the view specified by \viewSpec\ by reusing functional dependencies from \relationSet\   %
  and annotating each FD with its provenance information under the form of a provenance triple.} %
}

In order to compute the functional dependencies for a given view, one needs to tackle the complexity of the 
FD mining problem and thus to tame the size of explored FDs lattices. 

\textRev{In the following theorem, we state that at each computation step of a view without projection, the FDs previously mined continue to be valid. Conversely, we show that the application of a projection can only lead to the suppression of previously valid FDs, and never to the discovery of new FDs.} %

\begin{theorem}\label{thm:fdsConservation}\thmFont
    Let $\viewSpecArg{}$, $\viewSpecArg{1}$, and $\viewSpecArg{2}$ be view specifications.
    Let \fdSet, $\fdSet_1$ and $\fdSet_2$ be the sets of FDs over views specified by $\viewSpecArg{}$, $\viewSpecArg{1}$ and $\viewSpecArg{2}$, respectively.
    Let $\fds(V)$ denotes the set of FDs over a view specified by $V$. \\
    Then:
    \begin{align*}
        &\fds(\pi_X(\viewSpecArg{})) \subseteq \fdSet,  \qquad \fds(\sigma_\rho(\viewSpecArg{})) \supseteq \fdSet, \text{ and} \\[0.25cm]
        &\fds(\viewSpecArg{1} \diamond \viewSpecArg{2}) \supseteq \fdSet_1 \cup \fdSet_2 \text{, with }\diamond\in \{\bowtie;\fullouterjoin;\leftouterjoin;\rightouterjoin;\ltimes;\rtimes\}.
    \end{align*}
\end{theorem}

{\bf Our Solution. } 
In order to compute the set of FDs over integrated views with their provenance information, we propose the workflow illustrated in Figure~\ref{fig:workflow}.
It consists of three main steps as follows: %
(1) Mining of FDs over the base relations, limited to FDs with attributes projected in the resulting view (\jedi step 1); 
(2) Discovery of approximate single-table FDs that are upstaged and become exact FDs via the selection operations (\jedi step 2); 
(3) Discovery of all FDs that can appear via the join operations (\jedi step 3). \textRev{In other words, Step 2 allows to retrieve FDs that are approximate in the individual base tables but become exact due to the selection operations of the SPJ views, whereas Step 3 retrieves FDs due to join operations between multiple base tables including upstaged, inferred and join FDs.}

These steps and their corresponding algorithms are detailed in the next sections, respectively.
Their application is done by \jedi main algorithm (Algorithm~\ref{alg:mainFramework}) either during initialisation (lines\#\ref{line:main:beginMain}--\ref{line:main:endInstFDs} for Step 1)
or during the recursive traversal of the view specification tree (Subroutine \provFDs for Steps 2-3).

\subsection{Reducing Space of Explored Candidates using Projections}
As our goal is to discover the FDs occurring on top of an integrated view, we focus on mining FDs containing only attributes that are retrieved in the output  SPJ view. 

The number of considered attributes greatly influences the size of the explored lattices during FD mining. Thus, we make use of the knowledge of the projected attribute set at the very first step of our framework, in order to efficiently reduce the cost of mining FDs from the base relations. 
This can be seen in the main algorithm of our framework, Algorithm~\ref{alg:mainFramework}, during the mining of the FDs on the base relations  (lines \#~\ref{line:main:beginInstFDs}--\ref{line:main:endInstFDs}).
After this step, each following step of our framework relies on the FDs mined at the previous steps. Thus, additionally to reducing the FD discovery complexity over the base relations, removing unwanted FDs at this step also prevents useless computation during the following steps of our framework.

\begin{algorithm}[t]
\algoFontSize
\SetAlgoLined
\KwInput{\relationSet: a set of relational instances;\\ 
\hspace{1cm}\viewSpec: a view specification over instances in \relationSet. \\
}
\KwResult{the set of provenance triples of the FDs over \viewSpec.}

\BlankLine
\SetKwFunction{computeFDs}{provFDs}
\SetKwFunction{selectionFDs}{selectionFDs}
\SetKwFunction{upstagedFDs}{joinUpFDs}
\SetKwFunction{inferredFDs}{inferFDs}
\SetKwFunction{joinFDs}{joinFDs}

$\relationsFDsSets \gets \emptyset$\;\label{line:main:beginMain}
$\projectedAttsSet \gets $ \compute the set of attributes $\projectedAttributes(\viewSpec)$ \;

\For{each relation ${R_i}\in \relationSet$\label{line:main:beginInstFDs}}
{
    $\fdSet_{R_i} \gets$ \compute FDs in $R_i$ limited to attributes in \projectedAttsSet\;
    \add $\fdSet_{R_i}$ to \relationsFDsSets\;\label{line:main:endInstFDs}
}
\KwRet{$\computeFDs{\relationSet, \viewSpec, \relationsFDsSets, \projectedAttsSet}$}\;
\BlankLine
\SetKwProg{myproc}{Subroutine}{}{}
\myproc{\computeFDs{$\relationSet$, \viewSpec, $\relationsFDsSets$, $\projectedAttsSet$}}{
\Switch{\viewSpec}{
            \Case{$R \in \relationSet$\label{line:baseCase}}{
                $\provenanceSet\gets\emptyset$\;
                $\fdSet\gets$ \get $\fdSet_{R}$ in \relationsFDsSets\;
                \For{each FD $d \in\fdSet$}
                    {
                    \add triple $(d, \mathrm{"base"},R)$ to \provenanceSet
                    
                    }     
                \KwRet \provenanceSet\;
            }
            \Case{$\pi_X(V^\prime_{\relationSet})$\label{line:projectionCase}}{
                \KwRet $\computeFDs(\relationSet, V^\prime_{\relationSet}, \relationsFDsSets, \projectedAttsSet)$\;
            }
            \Case{$\sigma_\rho(V^\prime_{\relationSet})$\label{line:selectionCase}}{
                $\provenanceSet \gets  \computeFDs(\relationSet, V^\prime_{\relationSet}, \relationsFDsSets, \projectedAttsSet)$\;
                \KwRet $\provenanceSet \cup  \selectionFDs(\relationSet, V^\prime_{\relationSet}, \provenanceSet, \rho)$\;
            }
            \Case{$V^\prime_{\relationSet} \diamond V^{\prime\prime}_{\relationSet}$\label{line:joinCase}}{
                $\provenanceSet_{rec} \gets  \computeFDs(\relationSet, V^\prime_{\relationSet}, \relationsFDsSets, \projectedAttsSet)$\;
                $\provenanceSet_{rec} \gets  \computeFDs(\relationSet, V^{\prime\prime}_{\relationSet}, \relationsFDsSets, \projectedAttsSet)$\;
                $\provenanceSet_{up} \gets$ \upstagedFDs(\projectedAttsSet)\;
                $\provenanceSet_{inf} \gets$\inferredFDs()\;
                $\provenanceSet_{join} \gets$\joinFDs()\;
                \KwRet $\provenanceSet_{rec} \cup \provenanceSet_{up} \cup \provenanceSet_{inf} \cup \provenanceSet_{join}$ \;
            }
}
}
\caption{$\mathtt{InFine}$}
\label{alg:mainFramework}
\end{algorithm}
\subsection{Mining of FDs Through View Selections}
New FDs may appear due to selections in the view definition.
This will occur when the selection operation leads to filter the tuples violating an FD, making this FD valid in the resulting instance. In such a case, our framework aims at producing provenance triples for these new FDs with  Algorithm~\ref{alg:upstagedSelFDs}. This algorithm checks if the selection operation leads to some tuple filtering  (from line~\#\ref{line:algoSel:hasFiltering})  to avoid unnecessary mining step if the filter is not applicable. If some tuples are filtered by the selection, then upstaged FDs need to be mined and, to do so, we rely on a level-wise approach for FD mining.
\textRev{The intuition behind such an approach is to explore the possible sets of attributes involved in the candidate FDs, by organizing them into a lattice beginning from an infimum with FDs with singleton as \texttt{lhs} (line~\#\ref{line:algoSel:initFirstLevel}). 
Then, FD validity is tested and next level candidates are generated (line~\#\ref{line:algoSel:initNLevel}) while taking into account the FDs discovered at previous levels to prune the candidates FDs (lines~\#\ref{line:algoSel:pruneTane}--\ref{line:algoSel:pruneAlreadyKnown}).
The algorithm stops when no candidates are left, and then the resulting set of minimal FDs are labelled with their provenance triples, the latter being returned by the algorithm.
}

\begin{algorithm}[!h]
\algoFontSize
\SetAlgoLined
\KwInput{$\relationSet$: a set of relational instances;\\ 
\hspace{1cm}\viewSpec: a view specification over instances in \relationSet;\\ 
\hspace{1cm}$\rho$: the selection condition;\\
\hspace{1cm}$\provenanceSet_{\view}$: the sets of provenance triples of \viewSpec;\\
\hspace{1cm}\projectedAttsSet: the set of projected attributes to explore.}
\KwResult{upstaged FDs provenance triples over \viewSpec}
\BlankLine
$\provenanceSet \gets \emptyset$\;
$\fdSet_{\view} \gets \get$ the set of FDs in the triples in $\provenanceSet_{\view}$\;
 $\view_{sel} \gets \sigma_\rho(\viewSpec)$\;
\uIf{$\fun{size}(\view_{sel}) < \fun{size}(\viewSpec)$ \label{line:algoSel:hasFiltering}}
    {
    $\fdSet_{cand} \gets$ generate candidate FDs for first level of $\viewSpec$\;\label{line:algoSel:initFirstLevel}
    \prune FDs in $\fdSet_{sel}$ with attributes not in \projectedAttsSet\;
    \Repeat{$\fdSet_{cand} = \emptyset$}
    {
         \textRev{\prune non-minimal FDs in $\fdSet_{cand}$ knowing $\fdSet_{out}$} \;\label{line:algoSel:pruneTane}
         \textRev{\prune non-minimal FDs in $\fdSet_{cand}$ knowing $\fdSet_{\view}$} \;\label{line:algoSel:pruneAlreadyKnown}
         \add to $\fdSet_{out}$ the FDs from $\fdSet_{cand}$ holding in $V$\;
        $\fdSet_{cand} \gets$ generate candidate FDs for next level\;\label{line:algoSel:initNLevel}
    }
    \For{each FD $d \in\fdSet_{out}$}
    {
        \add triple $(d, \mathrm{"upstaged\ selection"},\viewSpec)$ to \provenanceSet
    }   
    \KwRet $\provenanceSet$
}
\Else{\KwRet{$\emptyset$}\;}

\caption{$\mathtt{selectionFDs}$} %
\label{alg:upstagedSelFDs}
\end{algorithm}

\subsection{Mining of FDs from Joined Tables}
\textRev{In the next sections, we will show how we can mine the upstaged, inferred and join FDs generated during a join operation. 
In the next theorem, we state that the join order can only affect the type of upstaged FDs which can switch between left or right upstaged FDs:
\begin{lemma}[FD types preservation though join operations]
Let $\relationSet = \{R_1;\dots;R_n\}$ be a set of relational instances.
Let $\fds(R_i)$ denotes the set of minimal FDs over the relational instance $R_i$.
Let $\Sigma_{left\_up},\Sigma_{right\_up}, \Sigma_{inf} \text{ and } \Sigma_{join}$ denote the sets of left upstaged, right upstaged, inferred and join FDs over $R_1 \diamond\dots\diamond R_n$, respectively.
Then, the sets $\Sigma_{left\_up}\cup\Sigma_{right\_up}, \Sigma_{inf} \text{ and } \Sigma_{join}$ are equal regardless of the join order.
At the opposite, equality between sets $\Sigma_{left\_up}$ and $\Sigma_{right\_up}$ cannot be guaranteed.

\end{lemma}

Next, we show how upstaged FDs can be mined over joins.
}
\subsubsection{Mining of Upstaged FDs from Joined Tables}
Upstaged FDs may appear due to the join operations in the SPJ view when tuples from one base relation cannot be joined with their counterpart in the other base relation, i.e., when some join attribute values are missing in one of the tables. %
This mechanism is expressed more formally in the following lemma:

\begin{lemma}[Upstaged join FDs]\thmFont
 Let $\leftInst$ and $\rightInst$ be two instances over relations $\leftSch$ and $\rightSch$, respectively, and $\fdSet_\leftInst$ and $\fdSet_\rightInst$ be the two sets of all FDs such that $\leftInst \models \fdSet_\leftInst$ and $\rightInst \models \fdSet_\rightInst$, respectively. \\ Then the sets of upstaged join FDs denoted $\fdSet^{new}_{{\leftInst}}$ and $\fdSet^{new}_{{\rightInst}}$ are the sets:
  \begin{align}
      \fdSet^{new}_{{\leftInst}} &= \{ d\ |\ d\not\in\fdSet_{\leftInst} \wedge \left(\leftInst \Diamond_{\leftJoinAtt=\rightJoinAtt} (\pi_{\rightJoinAtt}(\rightInst))\models d \right)\} \\
    \fdSet^{new}_{{\rightInst}} &= \{ d\ |\ d\not\in\fdSet_{\rightInst} \wedge \left(\pi_{\leftJoinAtt}(\leftInst) \Diamond_{\leftJoinAtt=\rightJoinAtt} \rightInst)\models d \right)\}
  \end{align}
\end{lemma}

\begin{example}
To illustrate the case of upstaged join FDs we use the example in Figure 1.
The FD ${\small\texttt{expire\_flag}}  \approxi{1} {\small\texttt{dod}}$ in Table {\small\texttt{PATIENT}} is violated by the presence of two tuples for patient (\#257). However, in the join result of ${\small\texttt{PATIENT}} \bowtie_{{\small\texttt{subject\_id}}} {\small\texttt{ADMISSION}}$, the violating tuple \#257 has no counterpart in the {\small\texttt{ADMISSION}} table and it disappears from the join result. Consequently, the FD  ${\small\texttt{expire\_flag}} \to {\small\texttt{dod}}$ becomes valid in the join result. 

\begin{algorithm}[!h]
\algoFontSize
\SetAlgoLined
\KwInput{$\leftView,\rightView$: two view specifications;\\ 
\hspace{1cm}$\leftJoinAtt,\rightJoinAtt$: the sets of join attributes for $\leftView$ and $\rightView$; \\ 
\hspace{1cm}$\Diamond$: a join operator in $\{\bowtie;\ltimes;\rtimes;\fullouterjoin;\leftouterjoin;\rightouterjoin\}$; \\
\hspace{1cm}$\provenanceSet_{\leftView},\provenanceSet_{\rightView}$: the provenance triples sets of $\leftView$ and $\rightView$;\\
\hspace{1cm}\projectedAttsSet: the set of projected attributes to explore.}
\KwResult{upstaged FDs provenance triples over \leftView and \rightView}

\BlankLine
\SetKwFunction{procFDs}{upstagedFDs}
\SetKwFunction{procAFDs}{upstagedAFDs}

$\provenanceSet \gets \emptyset$\;
$\fdSet_{\leftView} \gets \get$ the set of FDs in the triples in $\provenanceSet_{\leftView}$\;\label{line:joinUp:begin}
$\fdSet^{up}_{left} \gets$
    \procFDs{\leftView, \rightView, $X$, $Y$, $\fdSet_{\leftView}$,\projectedAttsSet}\; \label{line:joinUp:upFDsLeft}
\For{each FD $d \in\fdSet^{up}_{left}$}
    {
        \add triple $(d, \mathrm{"upstaged\ left"},\leftView\Diamond\rightView)$ to \provenanceSet\;
    }     
\BlankLine
$\fdSet_{\rightView} \gets \get$ the set of FDs in the triples in $\fdSet{\rightView}$\;
$\fdSet^{up}_{right} \gets$
    \procFDs{\rightView, \leftView, $Y$, $X$, $\fdSet_{\rightView}$,\projectedAttsSet}\; \label{line:joinUp:upFDsRight}
\For{each FD $d \in\fdSet^{up}_{right}$}
    {
        \add triple $(d, \mathrm{"upstaged\ right"},\leftView\Diamond\rightView)$ to \provenanceSet\;\label{line:joinUp:end}
    }   
\BlankLine
\KwRet{\provenanceSet}\;
\BlankLine

\SetKwProg{myproc}{Subroutine}{}{}
\myproc{\procFDs{$I$,$J$,$X$,$Y$,$\fdSet$,\projectedAttsSet}}{
$\fdSet_{out} \gets \emptyset$\;
 $I_{join} \gets I \Diamond_{X=Y} (\pi_{Y}(J))$\;\label{line:joinLeft}
\If{$\fun{size}(I_{join}) < \fun{size}(I)$\label{line:sizeLeft}} 
    {
    $\fdSet_{cand} \gets$ generate candidate FDs for first level of $I_{join}$\;
    \prune FDs in $\fdSet_{cand}$ with attributes not in \projectedAttsSet\;
    \Repeat{$\fdSet_{cand} = \emptyset$}
    {
         \hspace{-0.2cm}\textRev{\prune non-minimal FDs in $\fdSet_{cand}$ knowing $\fdSet_{out}$}\;\label{line:algoJoin:pruneTane}
         \hspace{-0.2cm}\textRev{\prune non-minimal FDs in $\fdSet_{cand}$ knowing $\fdSet_{\view}$}\;\label{line:algoJoin:pruneAlreadyKnown}
         \hspace{-0.2cm}\add to $\fdSet_{out}$ the FDs from $\fdSet_{cand}$ holding in $I$\;
        \hspace{-0.2cm}$\fdSet_{cand} \gets$ generate candidate FDs for next level\;
    }
}
\KwRet $\fdSet_{out}$}
\caption{\compUpstaged} %
\label{alg:filteringFDs}
\end{algorithm}

\end{example}

To compute upstaged FDs, we propose Algorithm~\ref{alg:filteringFDs}. Lines~\#\ref{line:joinUp:begin}--\ref{line:joinUp:end} handle the inputs of the SPJ view for each table participating in the join operation. 
For each side of the join, the subroutine \procFDs is executed (lines~\#\ref{line:joinUp:upFDsLeft} and \ref{line:joinUp:upFDsRight}) and computes partially the join only with the join attributes from the left side table (line~\#\ref{line:joinLeft}) to  check the assumption of the join value set preservation \cite{booksGarcia}. If the assumption is violated (i.e., if some tuples have been deleted through the join operation, as checked at line~\#\ref{line:sizeLeft}), some upstaged join FDs are produced. 
\textRev{The subroutine works analogously to Algorithm~\ref{alg:upstagedSelFDs} to discover the FDs in the input instance by using a level-wise approach where the previously discovered FDs are used to improves the pruning of candidates (lines~\#\ref{line:algoJoin:pruneTane}--\ref{line:algoJoin:pruneAlreadyKnown}).}

In this algorithm, the computation is performed over upstaged FDs only. %
Next, we discover the inferred %
FDs by relying on the characteristics of the join and on the FDs discovered previously.

\subsubsection{Mining of Inferred FDs from Joined Tables}
We will now show how the inferred FDs can be deduced from the sets of FDs discovered from the tables involved in the join operation. 
\textRev{In the next lemma, we show that if the right-hand side (\texttt{rhs}) of an FD is not functionally defined by the set of join attributes, then this FD cannot be an inferred FD in a join result. This property of the join operation is formally defined as follow:}

\begin{lemma}\label{lemma:notModelsImplication}\thmFont
Let $\leftInst$ and $\rightInst$ be two instances over relations $\leftSch$ and $\rightSch$, respectively. Let $\leftInst \Diamond_{\leftJoinAtt=\rightJoinAtt} \rightInst$ be a join result with $\leftJoinAtt \subseteq \atts(\leftInst)$, $\rightJoinAtt  \subseteq \atts(\rightInst)$. For all $A  \subseteq \atts(\leftInst)\setminus \leftJoinAtt$ and $B \subseteq \atts(\rightInst)\setminus \rightJoinAtt$:
$$\text{if }\leftInst \Diamond_{\leftJoinAtt=\rightJoinAtt} \rightInst \not\models \leftJoinAtt \to B \text{ then } \leftInst \Diamond_{\leftJoinAtt=\rightJoinAtt} \rightInst \not\models A \to B $$
\end{lemma}
\textRev{
\begin{example}
    To illustrate the property proved in Lemma~\ref{lemma:notModelsImplication}, we observe that the diagnosis is not determined by the patient identifier in Figure 1, for example patient \#249 has been admitted three times for a different pathology each time, i.e.,  ${\small\texttt{PATIENT}}\bowtie_{ {\small\texttt{subject\_id}}} {\small\texttt{ADMISSION}} \not\models  {\small\texttt{subject\_id}} \to  {\small\texttt{diagnosis}}$. 
    From Lemma~\ref{lemma:notModelsImplication}, we know that {\small\texttt{diagnosis}} in the join result ${\small\texttt{PATIENT}}\bowtie_{{\small\texttt{subject\_id}}} {\small\texttt{ADMISSION}}$ cannot be determined by any set of attributes coming from \texttt{PATIENT} table. Such similar inferences may be trivial for the user, but they usually require the knowledge of the attribute semantics. If not encoded, they are difficult  to capture by a system. However, the property shown in Lemma~\ref{lemma:notModelsImplication} can be used to drastically reduce the set of possible FDs that can appear after a join operation.
\end{example}
}
\textRev{Intuitively, from Lemma~\ref{lemma:notModelsImplication} follow that inferring FDs in the results of a join operation using Armstrong's transitivity axiom can only be done if the transitivity is done through the join attributes. This is formalized in the following theorem:}
\begin{theorem}\label{thm:MonoInstanceCrossJoinFDs}\thmFont
  Let $\leftInst$ and $\rightInst$ be two instances over relations $\leftSch$ and $\rightSch$, respectively. 
Let $\leftInst \Diamond_{\leftJoinAtt=\rightJoinAtt} \rightInst$ be a join result with $\leftJoinAtt \subseteq \atts(\leftInst)$, $\rightJoinAtt  \subseteq \atts(\rightInst)$. 
For all $A  \subseteq \atts(\leftInst)\setminus \leftJoinAtt$ and $B \subseteq \atts(\rightInst)\setminus \rightJoinAtt$,\\
 If $\leftInst \Diamond_{\leftJoinAtt=\rightJoinAtt} \rightInst \models A \to \leftJoinAtt \wedge  \leftInst \Diamond_{\leftJoinAtt=\rightJoinAtt} \rightInst \models \leftJoinAtt \to B$,\\
 Then $\leftInst \Diamond_{\leftJoinAtt=\rightJoinAtt} \rightInst \models A \to B$.
\end{theorem}

\textRev{
\begin{example}
In ${\small\texttt{PATIENT}} \bowtie_{{\small\texttt{ subject\_id}}}{\small\texttt{ADMISSION}}$ result illustrated in Figure 1, we observe that the diagnosis determines the date of birth, i.e., 
${\small\texttt{diagnosis}} \to {\small\texttt{dob}}.$
The reason is that we have:  
${\small\texttt{admission\_location}}$, ${\small\texttt{diagnosis}} \to {\small\texttt{subject\_id}}$ in ${\small\texttt{ADMISSION}}$  and ${\small\texttt{subject\_id}} \to {\small\texttt{dob}}$ in ${\small\texttt{PATIENT}}$.  
Since these tables do not contain any null values for the \texttt{lhs} and \texttt{rhs} attributes, joining them with attribute ${\small\texttt{subject\_id}}$ leaves these FDs unchanged with no violation. By transitivity, we obtain:
${\small\texttt{diagnosis}} \to {\small\texttt{dob}}.$
\end{example}
}
\begin{algorithm}[!h]
\algoFontSize
\SetAlgoLined
\KwInput{$\leftView,\rightView$: two view specifications;\\ 
\hspace{1cm}$\leftJoinAtt,\rightJoinAtt$: the sets of join attributes for $\leftView$ and $\rightView$; \\ 
\hspace{1cm}$\Diamond$: a join operator in $\{\bowtie;\ltimes;\rtimes;\fullouterjoin;\leftouterjoin;\rightouterjoin\}$; \\
\hspace{1cm}$\provenanceSet_{\leftView},\provenanceSet_{\rightView}$: the provenance triples sets of $\leftView$ and $\rightView$;\\
\hspace{1cm}\projectedAttsSet: the set of projected attributes to explore.}
\KwResult{inferred FDs provenance triples over ${\leftInst \Diamond_{\leftJoinAtt=\rightJoinAtt} \rightInst}$}
\BlankLine

\SetKwFunction{proc}{infer}
\SetKwFunction{procRef}{refine}
$\fdSet_{\leftView} \gets \get$ the set of FDs in the triples in $\provenanceSet_{\leftView}$\;
$\fdSet_{\rightView} \gets \get$ the set of FDs in the triples in $\provenanceSet_{\rightView}$\;
\BlankLine
$\fdSet_{infL} \gets$ \proc{$\leftJoinAtt$,$\rightJoinAtt$,$\fdSet_{{\leftView}}$,$\fdSet_{{\rightView}}$}\;
$\fdSet_{infR} \gets$ \proc{$\rightJoinAtt$,$\leftJoinAtt$,$\fdSet_{{\leftView}}$,$\fdSet_{{\rightView}}$}\;
$\fdSet_{inf} \gets \fdSet_{infL} \cup \fdSet_{infR}$\;
\BlankLine
$\fdSet_{ref} \gets$ \procRef{$\leftInst$,$\rightInst$,$X$,$Y$,$\fdSet_{inf}$, $\Diamond$}\;
\For{each FD $d \in\fdSet_{ref}$}
    {
        \add triple $(d, \mathrm{"inferred"},\leftView\Diamond\rightView)$ to \provenanceSet
    }     
\KwRet \provenanceSet\;
\BlankLine
\SetKwProg{myproc}{Subroutine}{}{}
\myproc{\proc{$X$,$Y$,$\fdSet$, $\fdSet^\prime$}}{
$\fdSet_{out} \gets \emptyset$\;
\ForAll{$A \to X$ in $\fdSet$\label{line:transX}}
{
    \ForAll{$Y \to b$ in $\fdSet^\prime$\label{line:transY}}
    {
       \add $A \to b$ to $\fdSet_{out}$\;
    }
}
\KwRet $\fdSet_{out}$\;}

\BlankLine
\SetKwProg{myproc}{Subroutine}{}{}
\myproc{\procRef{$\leftInst$,$\rightInst$,$X$,$Y$,$\fdSet_{inf}$, $\Diamond$}}{
Let $\fdSet_{out} \gets \fdSet_{inf}$\;
\ForAll{$A \to b$ in $\fdSet_{inf}$\label{line:fdRef}}
{
    Let $I \gets \pi_{X \cup A}(\leftInst) \Diamond \pi_{Y \cup \{b\}}(\rightInst)$\label{line:instRef}\;
    \ForAll{$A^\prime \subset A$\label{line:fdSub1}}
    {
        \If{$A^\prime \to b$ holds in $I$\label{line:testFdHolds}\label{line:fdSub2}}
        {
        \add $A^\prime \to b$ to $\fdSet_{out}$\;
        \prune non-minimal FDs in $\fdSet_{out}$ knowing that $A^\prime \to b$ is valid\;
        }
    }
}
 \KwRet $\fdSet_{out}$\;}
\caption{$\mathtt{inferFDs}$} %
\label{algo:MonoInst_CrossJoinFDs}
\end{algorithm}

To compute the set of FDs with \texttt{lhs} attributes coming from a single instance, as described in Theorem~\ref{thm:MonoInstanceCrossJoinFDs}, we propose Algorithm~\ref{algo:MonoInst_CrossJoinFDs}.
First, the subroutine \subInfer extracts the FDs that can be retrieved by transitivity (lines~\#\ref{line:transX} and~\ref{line:transY} in subroutine \subInfer). Note that in the case of equijoins, equality of values might be enforced between sets of attributes with different names (i.e., $X$ and $Y$ might be different), thus for the general case, the FD (line~\#\ref{line:transY}) cannot be simplified into an FD $X\to b$. Conversely, if we restrict the join operations to natural joins only, such a simplification can be made.

Then, for each FD returned by \subInfer, 
the subroutine \subRefine checks whether the FD is minimal or if a subset of its \texttt{lhs} leads to a minimal FD.
To do so, subroutine \subRefine uses an horizontal partition of the joined instances in which only the necessary attributes to perform the verification are considered (line~\#\ref{line:instRef}).
These necessary attributes are the join attributes (to perform the join operation), and the \texttt{lhs} and \texttt{rhs} attributes of the refined FD $A \to b$ (line~\#\ref{line:fdRef}) as \subRefine only considers candidates with subsets of $A$ as \texttt{lhs} and $b$ as \texttt{rhs} (lines~\#\ref{line:fdSub1} and \ref{line:fdSub2}).

\subsubsection{Mining of Join FDs from Joined Tables}

\textRev{Now, we characterize the set of join FDs which hold on a join result.
Contrarily to the inferred FDs that can be deduced directly using a simple logical reasoning, these join FDs need to be discovered and validated from the data. }
For example, {\small\texttt{gender, expire\_flag}}$ \to $ {\small\texttt{insurance}} of our  example has attributes from  {\small\texttt{PATIENT}} in \texttt{lhs} and attributes from  {\small\texttt{ADMISSION}} in \texttt{rhs} and it cannot be inferred logically. Other FDs with the same properties are illustrated in orange in Figure ~1. In the following theorem, we show that if \texttt{lhs} attributes of an FD come from the instances participating in the join, then we cannot predict their validity without checking them directly with some representative (if not all) tuples of the join result:

\begin{theorem}\thmFont\label{thm:infNotSufficient}
Let $\leftInst$ and $\rightInst$ be two instances over relations $\leftSch$ and $\rightSch$, respectively. Let $\leftInst \Diamond_{\leftJoinAtt=\rightJoinAtt} \rightInst$ be a join result with $\leftJoinAtt \subseteq \atts(\leftInst)$, $\rightJoinAtt  \subseteq \atts(\rightInst)$. 
We cannot guarantee that all FDs over $\leftInst \Diamond_{\leftJoinAtt=\rightJoinAtt} \rightInst$ can be inferred from Armstrong's axioms over the FDs over $\leftInst$ and $\rightInst$ taken separately.
\end{theorem}
\textRev{
Such FDs are illustrated in the following example:
 \begin{example}\label{ex:joinFD}In our example of Figure~\ref{fig:example}, FDs that cannot be inferred are highlighted in orange.
For example,  {\small\texttt{gender, h\_expire\_flag}}$\to$ {\small\texttt{insurance}} is specific to the join of {\small\texttt{PATIENT}} and {\small\texttt{ADMISSION}}. It holds in {\small\texttt{PATIENT}} $\bowtie$ {\small\texttt{ADMISSION}}. Attributes {\small\texttt{gender}} and {\small\texttt{expire\_flag}} come from  {\small\texttt{PATIENT}} and attribute {\small\texttt{insurance}} comes from {\small\texttt{ADMISSION}}.
\end{example}
}
Theorem~\ref{thm:infNotSufficient} motivates the need for designing a new method for computing FDs from partial join results, as we cannot always infer all the FDs only using logical reasoning. However, we can rely on the following theorem to greatly reduce number of remaining FDs to check from the data:

\begin{theorem}\label{theo6}\thmFont
 Let $\leftInst$ and $\rightInst$ be two instances over relations $\leftSch$ and $\rightSch$, respectively. 
Let $\leftInst \Diamond_{\leftJoinAtt=\rightJoinAtt} \rightInst$ be a join result with $\leftJoinAtt \subseteq \atts(\leftInst)$, $\rightJoinAtt  \subseteq \atts(\rightInst)$. 
For all $A  \subseteq \atts(\leftInst)$, $A^\prime  \subseteq \atts(\rightInst)$ and $b \in \atts(\rightInst)$:
If $\leftInst \Diamond_{\leftJoinAtt=\rightJoinAtt} \rightInst \models AA^\prime \to b$, Then $\leftInst \Diamond_{\leftJoinAtt=\rightJoinAtt} \rightInst \models \rightJoinAtt A^\prime \to b$.
\end{theorem}

\begin{algorithm}[t]
\algoFontSize
\SetAlgoLined
\SetKwFunction{proc}{mine}
\KwInput{$\leftView,\rightView$: two view specifications;\\ 
\hspace{1cm}$\leftJoinAtt,\rightJoinAtt$: the sets of join attributes for $\leftView$ and $\rightView$; \\ 
\hspace{1cm}$\Diamond$: a join operator in $\{\bowtie;\ltimes;\rtimes;\fullouterjoin;\leftouterjoin;\rightouterjoin\}$; \\
\hspace{1cm}$\provenanceSet_{\leftView},\provenanceSet_{\rightView}$: the provenance triples sets of $\leftView$ and $\rightView$;\\
\hspace{1cm}$\provenanceSet^{inf}_{\leftView\Diamond\rightView}$: the inferred provenance triples set.}
\KwResult{join FDs provenance triples over ${\leftInst \Diamond_{\leftJoinAtt=\rightJoinAtt} \rightInst}$}

$\fdSet_{inf} \gets \get$ the set of FDs in the triples in $\provenanceSet^{inf}_{\leftView\Diamond\rightView}$\;
\BlankLine
$\fdSet_{\leftView} \gets \get$ the set of FDs in the triples in $\provenanceSet_{\leftView}$\;
$\fdSet_{left} \gets$ \proc{$\leftInst$,$\rightInst$,$\leftJoinAtt$,$\rightJoinAtt$,$\fdSet_{\rightView}$,$\fdSet_{inf}$}\;
\BlankLine
$\fdSet_{\rightView} \gets \get$ the set of FDs in the triples in $\provenanceSet_{\rightView}$\;
$\fdSet_{right} \gets$ \proc{$\rightInst$,$\leftInst$,$\rightJoinAtt$,$\leftJoinAtt$,$\fdSet_{\leftView}$,$\fdSet_{inf}$}\;
\BlankLine
\For{each FD $d \in\fdSet_{left} \cup \fdSet_{right}$}
    {
        \add triple $(d, \mathrm{"join FD"},\leftView\Diamond\rightView)$ to \provenanceSet
    }     
\KwRet \provenanceSet\;
\BlankLine
\SetKwProg{myproc}{Subroutine}{}{}
\myproc{\proc{$I$,$J$,$X$,$Y$,$\fdSet_{J}$,$\fdSet_{inf}$}}{
$\fdSet_{out} \gets \emptyset$\;
\ForAll{$Y \to b$ in $\fdSet_{{J}}$\label{line:joinAtt}}
{
    \ForAll{$A\subseteq atts(I)\setminus X$ \label{line:joinAttLhs}}
    {
    \If{$\not\exists A^\prime \subset A, A^\prime \to b \in \fdSet_{inf}$ and $A \to b$ holds in $I \Diamond J$\label{line:joinAttLhsEnd}}
            {
                add $A \to b$ to $\fdSet_{out}$\;
            }
    }
}
\ForAll{$YA^\prime \to b \in \fdSet_{J}$ such that $A^\prime\not\to b$\label{line:joinAttAndOthers}}
    {
        \ForAll{$A \cup A^\prime \to b$ such that $A \subseteq atts(I)\setminus X$\label{line:joinAttAndOthersLhs}}
        {
            \If{$A \cup A^\prime \to b$ holds in $I \Diamond J$\label{line:joinAttAndOthersLhsEnd}}
            {
                add $A \cup A^\prime \to b$ to $\fdSet_{out}$\;
            }
        }
    }
\KwRet $\fdSet_{out}$\;
}
\caption{$\mathtt{mineFDs}$} %
\label{algo:MultiInst_CrossJoinFDs}
\end{algorithm}

In-line with Theorem~\ref{theo6}, we propose Algorithm~\ref{algo:MultiInst_CrossJoinFDs} for selective mining and use the FDs previously discovered with Algorithms~\ref{alg:filteringFDs} and \ref{algo:MonoInst_CrossJoinFDs} to compute the remaining join FDs. Intuitively, Theorem~\ref{theo6} shows that a given attribute $b$ can be a \texttt{rhs} of a remaining join FDs only if we have previously found an FD of the form $YA \to b$ with $Y$ being the join attributes of the instance containing $b$.
Thus, it allows us to focus only on the plausible \texttt{rhs} (lines~\#\ref{line:joinAtt} and \ref{line:joinAttAndOthers} in subroutine \texttt{discover}) and explore their candidate \texttt{lhs} (lines~\#\ref{line:joinAttLhs}-\ref{line:joinAttLhsEnd} and \ref{line:joinAttAndOthersLhs}-\ref{line:joinAttAndOthersLhsEnd}).  In practice, there is no need to generate every candidate FDs initially. Instead, candidate FDs can be explored by generating a first level containing only the smallest candidates and by generating upper levels only when currently evaluated candidates are not valid. Moreover, we can avoid the computation of the full join by deleting a given \texttt{lhs} attribute $a$ if $a$ is not a possible \texttt{rhs} and, for every FD candidate $d: A\to b$ such that $a\in A$, $d$ is logically implied by previously discovered FDs.

\subsection{Completeness, Correctness, and Complexity}

In this section we show that the set of FDs retrieved by \jedi is both complete and correct, and present our solution complexity.
First, we show that \jedi always retrieve every functional dependency in the specified SPJ view.
\begin{theorem}[\jedi completeness]\thmFont
Let $\relationSet = \{R_1;\dots;R_n\}$ be a set of relational instances.
Let $\viewSpec$ be a view specification over relations in $\mathbf{R}$.
Let $\fds(\viewSpec)$ denotes the set of minimal FDs over the view specified by $\viewSpec$.
Let $\fds(\jedi)$ denotes the set of FDs computed by \jedi over the view specified by $\viewSpec$. Then: $$\forall d \in \fds(\viewSpec), \exists d^\prime \in \fds(\jedi) \text{ s.t. } d \equiv d^\prime$$
\end{theorem}

This establish the completeness of our approach. Now, in the next theorem, we show that every FD retrieved by \jedi is valid in the specified SPJ view: 
\begin{theorem}[\jedi correctness]\thmFont
Let $\relationSet = \{R_1;\dots;R_n\}$ be a set of relational instances.
Let $\viewSpec$ be a view specification over relations in $\mathbf{R}$.
Let $I_\viewSpec$ be the view specified by $\viewSpec$.
Let $\fds(\jedi)$ denotes the set of FDs computed by \jedi over the view specified by $\viewSpec$. Then: $$\forall d \in \fds(\jedi), I_\viewSpec \models d$$
\end{theorem}

We now detail the complexity of our algorithms.
 Algorithms~\ref{alg:upstagedSelFDs} and \ref{alg:filteringFDs} are based on level-wise algorithms through the attributes lattices. Their complexity is exponential in the number of attributes of the considered table. They prune candidates at each level when it is possible. In terms of memory, only two levels are required. The memory size is bounded by $\mathcal{O}\binom{k}{k/2}$ where $k$ is the number of attributes.
Algorithm~\ref{algo:MonoInst_CrossJoinFDs} infers and refines FDs coming from the previous step with complexity $\mathcal{O}(n \cdot f)$, where $f$ is the number of validated FDs and $n$ the maximal number of tuples in the left or right instance. 
The complexity of Algorithm~\ref{algo:MultiInst_CrossJoinFDs} is $\mathcal{O}(f \cdot f_j)$ where $f$ is the maximal number of validated FDs in the left or right instance and $f_j$ the number of validated FDs in the join instance. %

\section{Experiments}
\label{sec:experiments}

{\bf Evaluation Goals.} The two main points we seek to validate in our experimental study are as follows: (1) Does our approach enable us to discover all FDs \textRev{ with their provenance information} in an efficient manner and faster than the straightforward approach? (2) What is the impact of different data and SPJ view characteristics on the \jedi performance? 

{\bf Setup.} We perform all experiments on a laptop HP ZBook 15 machine with an Intel Core i7-4900MQ, 2.8 GHz, 32 GB RAM, powered by Windows 10 pro 64-bit. 
Our implementation in C++ uses only one thread. Sharable datasets, SPJ queries, scripts, and code are available at \url{https://github.com/ucomignani/InFine}. %

\begin{table}[t]
\centering
\small
\begin{tabular}{|c|l|r|r|}
\hline
{\bf DB} & {\bf Table}& {\bf (Att\# ; Tuple\#) }&{\bf FD\#}\\
\hline
\hline
 &{\bf Patients}& (7 ; 46.52k)& 11 \\
MIMIC3& {\bf Admissions}& (18 ; 58.976k)& 631\\
&{\bf Diagnoses\_icd}& (4 ; 651.047k)  & 2\\
&{\bf D\_icd\_Diagnoses}& (3 ; 14.710k)& 2\\
\hline
&{\bf active}& (2; 300)& 1\\
PTE&{\bf bond}&(4 ; 9.317k)& 3\\
&{\bf atm}& (5 ; 9.189k)& 5\\
&{\bf drug}& (1 ; 340)& 0\\
\hline
&{\bf atom}& (3; 12.333k)& 2\\
PTC&{\bf connected}&(3 ; 24.758k)& 3\\
&{\bf bond}& (3 ; 12.379k)& 2\\
&{\bf molecule}& (2 ; 343)& 1\\
\hline
&{\bf Supplier}& (7 ; 10k)& 34 \\
TPC-H&{\bf Customer}&(8 ; 150k)& 51\\
&{\bf Orders}& (9 ; 1.5M)& 53 \\
&{\bf LineItem}& (16 ; 6M)& 3946 \\
&{\bf Nation}& (4 ; 23)& 9 \\
&{\bf Region}& (3 ; 5)& 6 \\
&{\bf Part}& (7 ; 200k)& 99 \\
&{\bf Partsupp}& (5 ; 800k)&11 \\
\hline
\end{tabular}
\caption{Data characteristics.}\label{exp:tables}
\end{table}

\begin{table}[t]
\centering
\small
\begin{tabular}{|c|p{4.8cm}|r|r|}
\hline
{\bf DB} & {\bf SPJ View}&   {\bf Tuple\#} & {\bf FD\#}\\
\hline
\hline

 \parbox[t]{2mm}{\multirow{4}{*}{\rotatebox[origin=c]{90}{{\bf MIMIC3}}}}&Q(patients $\bowtie$  admissions) &58,976 & 16\\
&	diagnosesicd $\bowtie$  patients&58,798&12\\
&	dicddiagnoses $\bowtie$  diagnosesicd	&658,498&22\\
&	[diagnosesicd$\bowtie$patients]$\bowtie$ dicddiagnoses&	658,498&44\\
\hline

 \parbox[t]{2mm}{\multirow{4}{*}{\rotatebox[origin=c]{90}{{\bf PTC}}}}
&	atom $\bowtie$  molecule&	9,111&4\\
&connected $\bowtie$  bond&24,758&8\\
&	[connected $\bowtie$  bond] $\bowtie$  molecule	&18,312&12\\
&	connected $\bowtie_{id1}$ [atom $\bowtie$  molecule]&	18,312&12\\

\hline

 \parbox[t]{2mm}{\multirow{4}{*}{\rotatebox[origin=c]{90}{{\bf PTE}}}}	&	atm $\bowtie$  drug&	9,189&5\\
&active $\bowtie$  drug&299&1\\
&	[bond $\bowtie$  drug] $\bowtie$  active&7,994&6\\

&	[atm $\bowtie$  bond $\bowtie$  atm] $\bowtie$  drug&	9,317&24\\
\hline

 \parbox[t]{2mm}{\multirow{4}{*}{\rotatebox[origin=c]{90}{{\bf TPC-H}}}}
& Q2$^*$(P $\bowtie$ PS $\bowtie$ S $\bowtie$ N $\bowtie$ R)  & 	21,696& 69\\
& Q3$^*$(C $\bowtie$ O $\bowtie$ L) &60,150&14 \\

& Q9$^*$(P $\bowtie$ PS $\bowtie$ S $\bowtie$ L $\bowtie$ O $\bowtie$ N)   &3,735,632&8 \\
& Q11$^*$(P  $\bowtie$ S $\bowtie$ N)  
&284,160&151 \\
\hline
\end{tabular}
\caption{SPJ queries considered in our experiments.
}\label{exp:spj}
\end{table}
 
{\bf Data.}
We use  three  real-world  datasets  and  one  synthetic dataset in our experiments: (1) MIMIC-3, a clinical database \footnote{\url{https://physionet.org/content/mimiciii/1.4/}} \cite{mimiciii}; 
 (2) PTE\footnote{\url{https://relational.fit.cvut.cz/dataset/PTE}}, a database for predictive toxicology evaluation, used to predict whether a compound is carcinogenic, and (3) PTC\footnote{\url{https://relational.fit.cvut.cz/dataset/PTC}}, the dataset from the Predictive Toxicology Challenge that consists of more than three hundreds of organic molecules marked according to their carcinogenicity on male and female mice and rats; and (4) the TPC-H Benchmark\footnote{\url{http://www.tpc.org/tpch/}} with scale-factor 1. 
The datasets characteristics are given in Table I %
 and the characteristics of the queries corresponding to the views on the above datasets are provided in Table II%
 . 
\textRev{To the best of our knowledge, all the datasets used in the literature for benchmarking FD discovery methods consist of single tables, while we use multi-table scenarios. Note that the numbers of discovered FDs in the literature datasets are not comparable as we do not employ the same datasets due to the SPJ views studied in our approach.} To handle multiple tables with SPJ queries, we propose our own benchmark datasets along with views on top of them. We adapted TPC-H queries by removing group-by and order-by statements (that are not addressed in our approach) and used the specified  constants\footnote{see query validation sections of the TPC documentation at \url{http://tpc.org/tpc_documents_current_versions/pdf/tpc-h_v3.0.0.pdf}}. SPJ queries for the other datasets were manually crafted to make use of multiple tables  and obtain joins of various sizes (from 2 to 6 tables),  number of tuples (from 299 to 3 millions), and coverage values (from 0.12 to  25,812.67) to show the performances of our method comparatively in a large and representative range of SPJ views.

We observed that the cardinalities and overlap of the join attribute values in the datasets are rarely preserved through a join operation and this has a great impact on FD discovery from SPJ queries depending on the join operator used. To quantify this phenomenon, we define a measure called coverage as follows:
\begin{align*}
   Coverage(R\Diamond L) =
       \frac{1}{2}  \bigg( &Cov(R\Diamond L, L, X) + Cov(R\Diamond L, R, Y)\bigg)
    \\
    \text{with }Cov(Join, I, a) &= \frac{1} {|\pi_{a}(I) |} \sum\limits_{\forall v \in \pi_{a}(I)} 
                \frac{|\sigma_{a=v}(Join)) |}
                     {|\sigma_{a=v}(I) |}.
\end{align*}

  where $X$ and $Y$ denote the join attributes of $\leftInst$ and $\rightInst$ base tables respectively. $I$ is a considered instance and $a$ the considered join attribute. If $Coverage(R\Diamond L)=0$, no tuple from $L$ can be joined with tuples in $R$. For $Coverage(R\Diamond L)<1$, some tuples in $L$ (or $R$) may be missing from the join result, as it is the case for patients \#257 in {\small\texttt{PATIENT}} and \#247, \#248, and \#253 in {\small\texttt{ADMISSION}} that do not have their counterparts in the other table in our example. For $Coverage(R\Diamond L)=1$, there are as many tuples in both tables $L$ and $R$ as in the join result. For $Coverage(R\Diamond L)>1$, there are more tuples in the join result than in tables $L$ or $R$ as some tuples may be repeated through the join. The coverage values of the corresponding views are given in Table III.

\begin{figure*}[!t]
	\centering
            {\includegraphics[width=\linewidth]{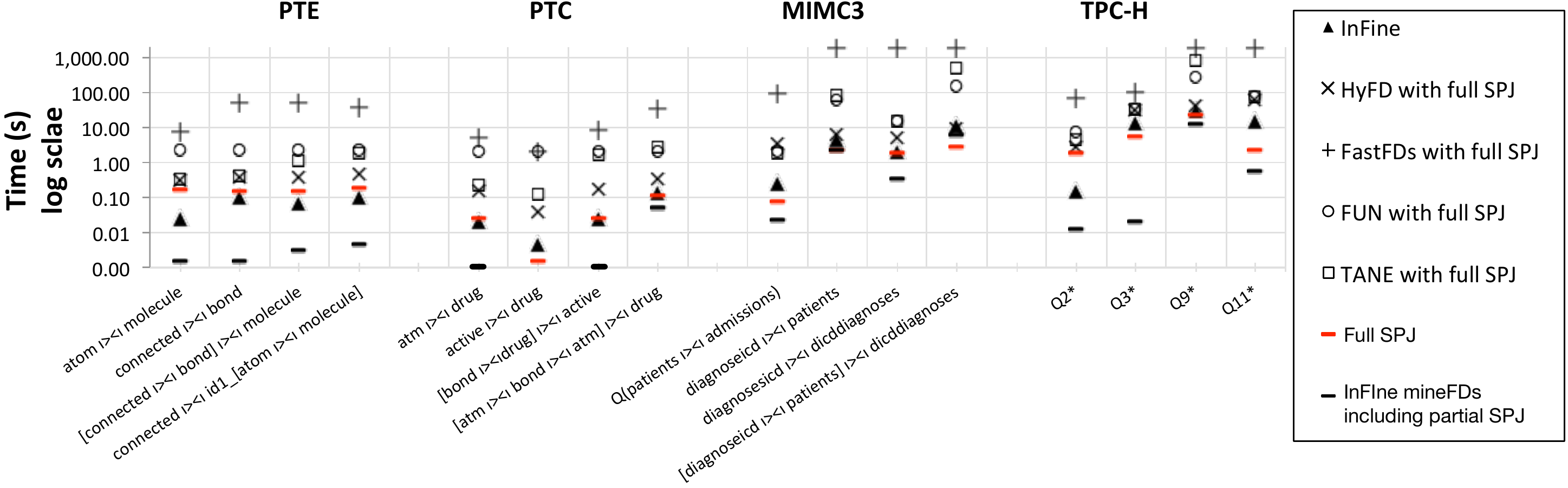}
            }
	\caption{Average runtime (in seconds):  \jedi against HyFD, FastFDs, FUN, and TANE \textRev{with full and partial SPJ computation}  }\label{fig:jedi-time-comparison}
\end{figure*} 

 \begin{figure*}[t]
	\centering
	\includegraphics[width=\linewidth]{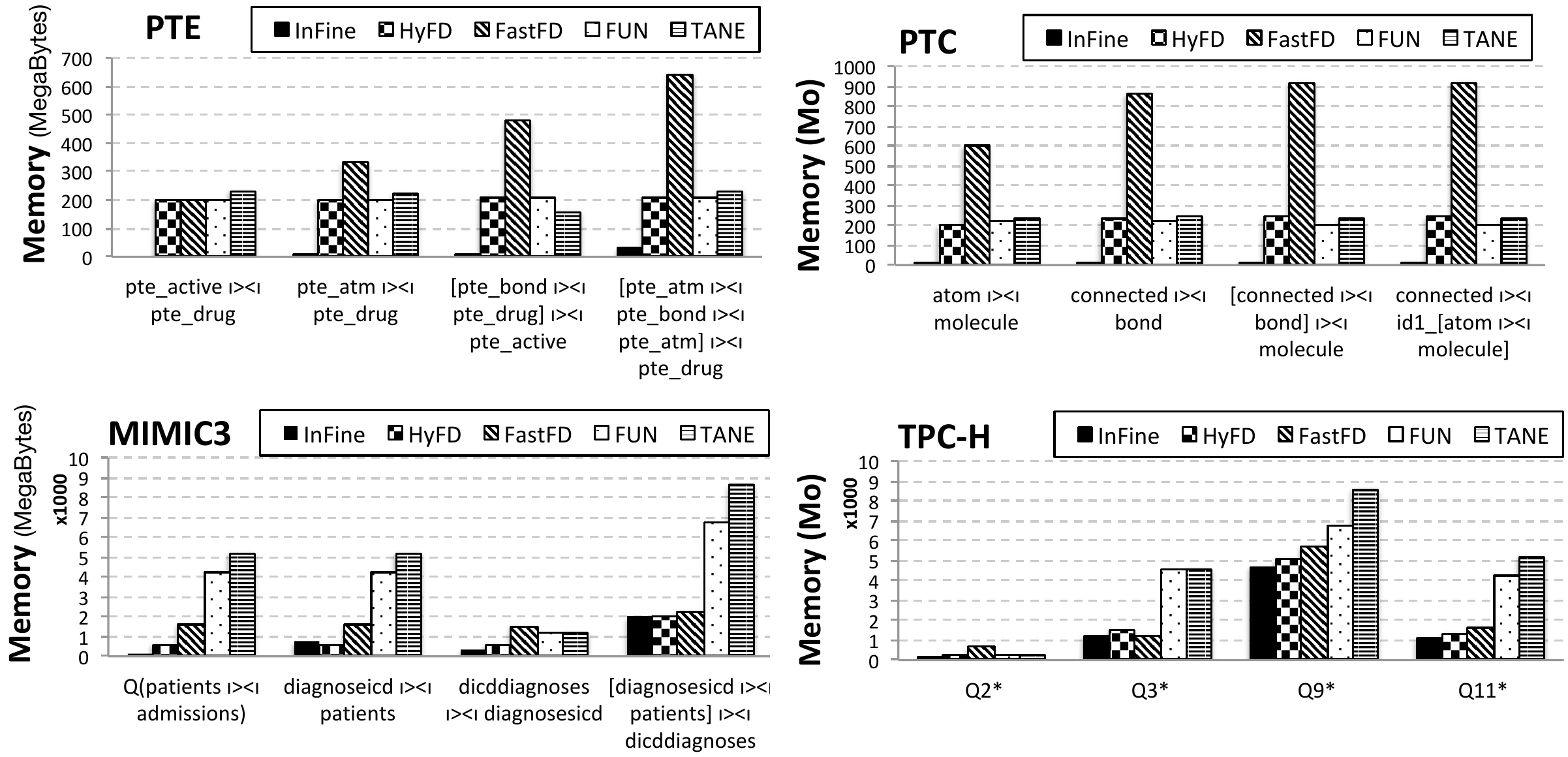}
	\caption{Maximal memory consumption (in \textRev{MegaBytes}) of \jedi against HyFD, FastFDs, FUN, and TANE    algorithms}\label{fig:jedi-memory}
\end{figure*}

{\bf Methods.} We compare the \jedi algorithms applied to base tables  
against four state-of-the-art FD discovery methods, i.e.: (1) TANE \cite{HKPT98,HKPT99},   (2) Fast\_FDs \cite{WGR01}, and (3) FUN \cite{NoCi01icdt}, and (4) HyFD \cite{HyFD} with Java implementation of Metanome \cite{Papenbrock:2015} using command line. 
The datasets employed in our study are stored in a PostgresSQL DBMS (version 12.7). Join attributes are indexed with both B-Tree and hash indexes. The reported results  correspond to the average over 10 runs for each query.

{\bf Metrics.}  
 We compute accuracy, average runtime, and  maximal memory consumption for each view. \textRev{In our setting, accuracy corresponds to precision and is defined as  the fraction of the number of FDs correctly discovered by  InFine applied only to the base tables over the total number of FDs found by baseline methods classically applied to the SPJ view results. Accuracy values are given in Table III and represented as percentages in pie charts of Fig. 5. All methods (including ours) reach accuracy of 1 at the end of their execution}. %

\textRev{ {\bf Comparison Setup.} Our goal is to compare our method which discovers FDs from a SPJ view specification and base tables and  provides FD provenance information against the straightforward approach which consists of discovering all the FDs from each base table, computing the view result, and discovering the FDs on top of the latter. Classical methods do not provide provenance information. Then, a fair comparison requires the preservation of FD provenance information from the base tables to the view.  To know the origin of each FD, both the baselines and our approach have to discover FDs from the base tables first. Since these execution times are the same in both cases, we don't include them. On the one hand, the competing methods are applied to each SPJ view result and we report their average execution time to which we added the execution time of the full SPJ view computation. On the other hand, \jedi is applied to the base tables only and it computes a partial SPJ view depending on the view specification (query sub-tree) and generate FD provenance triples; we report \jedi average execution time with time breakdown per algorithm. The time of the partial SPJ view computation is included in \texttt{mineFDs}. The generation of provenance triples is included in the execution of each \jedi algorithm in charge of annotating each FD accordingly. Total I/O times of \jedi are also  reported in Table III.}

\subsection{Efficiency Evaluation}

In a first set of experiments, we evaluate the runtime and memory consumption of the \jedi algorithms compared to the state-of-the-art FD discovery methods that follow the straightforward approach over the 16 SPJ queries on the real-world and synthetic datasets with a wide range of coverage values.  The selected queries are representative in terms of coverage and size (number of tables, tuples, and attributes). Results for other queries follow the same trend and are available as supplementary material on Github.

 \begin{figure*}[t]
	\centering
            {\includegraphics[width=.99\linewidth]{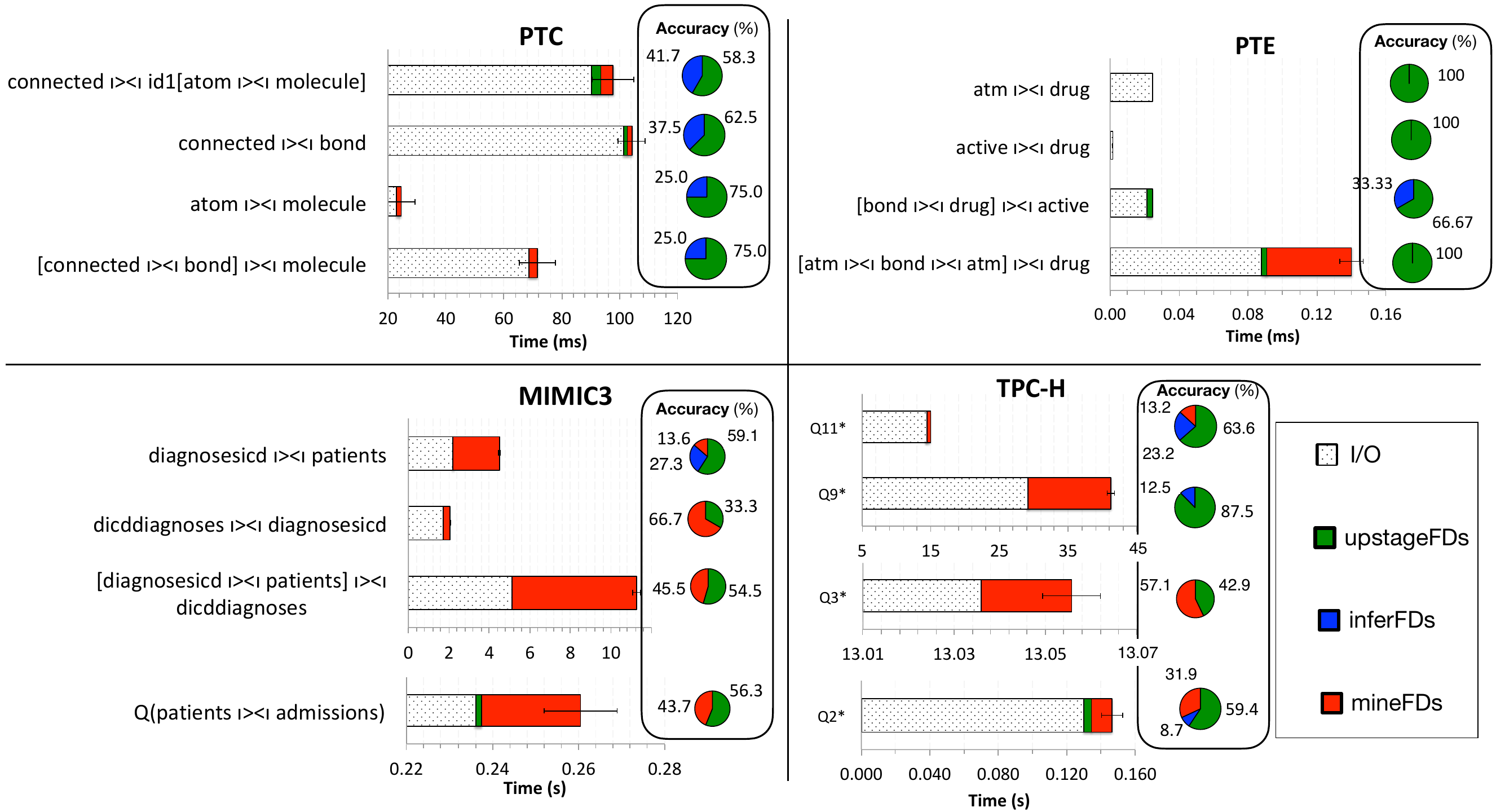}}

	\caption{Average runtime \textRev{and accuracy} of \jedi with breakdown per algorithm  
	}
	\label{fig:jedi-time-repartition}
\end{figure*}

\subsubsection{Runtime} 

Fig.~\ref{fig:jedi-time-comparison} presents,  for each method, the average total runtime in seconds for FD discovery (including data loading) in log scale. For the competing methods, we included the average execution time of the SPJ view over the indexed data. \textRev{ Similarly, we included the partial SPJ view computation time included in \texttt{mineFDs}. For all methods including ours, we did not include the time for discovering FDs from base tables since these costs are the same}.  Since our method does not require the full SPJ view computation to discover FDs and does not operate over the query result but only on the single base tables, it  is much faster than the traditional methods with one order of magnitude on average up to two orders of magnitude compared to Fast\_FDs ($> 2,000$ seconds).  \textRev{A  provenance triple is generated each time an FD is validated by one of the \jedi algorithms and is appended to the corresponding FD data structures with negligible time (included in each algorithm execution time).}
For MIMIC3 and TPC-H SPJ queries with the highest values for coverage and number of tuples (e.g., Q9$^*$ and Q11$^*$), \jedi is still outperforming the other methods. \textRev{The time difference between full and partial SPJ computations shown in Fig. 3 also explains the advantage of our approach.} %

\subsubsection{Memory Consumption} As shown in  Fig.~\ref{fig:jedi-memory}, the average maximal memory consumption of \jedi (for accuracy equals 1) is the lowest for all queries across the datasets. HyFD has the second position in terms of memory consumption efficiency and TANE is the worst one. \textRev{For all methods, Q9$^*$  has the highest memory consumption  due to its large size of 3.73 millions of tuples, very high coverage (25.8k), and the number of joins between 6 tables. This case is an example of a worst case scenario where many tuples are repeated through the join of the base tables. FD discovery from Q9$^*$  requires the mining and checking of all the valid FDs and storing all the relevant information. In this particular case,  \jedi time performance becomes comparable to HyFD over the full SJP view computation (due to \texttt{mineFDs}), although all FDs were discovered before by \texttt{upstageFDs} and \texttt{inferFDs} (see Fig. 5)}. %

\subsection{Quality Evaluation}
 \begin{table*}[t]
\centering
\scriptsize
\begin{tabular}{|c|l|p{1.6cm}|p{0.6cm}|p{1.1cm}|p{1cm}|p{1cm}|p{1.3cm}|r|p{1cm}|p{0.8cm}|}
\hline
{\bf DB} & {\bf SPJ View}&{\bf (Att\# ; Tuple\#)}& {\bf Cov. } & {\bf UpstageFDs \textRev{Accuracy}} & {\bf InferFDs \textRev{Accuracy} }& {\bf MineFDs \textRev{Accuracy}}& {\bf Total \textRev{Accuracy} (FD\#)}& {\bf I/O (s)}&{\bf upstageFDs (s)}&{\bf mineFDs (s)}\\
\hline
 \parbox[t]{2mm}{\multirow{4}{*}{\rotatebox[origin=c]{90}{{\bf PTE}}}}	&	atm $\bowtie$  drug&(5 ; 9,189)&	14.01& 1	&	0	&	0	& 1	(5 FDs) & 0.0246 &0.0000&0.0000\\
&active $\bowtie$  drug&(2 : 299)&	0.94& 	1	&	0	&	0	&	1 (1 FD)&0.0015&0.0000&0.0000\\
 
&	[bond $\bowtie$  drug] $\bowtie$  active&(6 ; 7,994)&	13.83	&	0.67	&	0.33	&	0	& 1	(6 FDs)& 0.0215&0.0030&0.0000\\

&	[atm $\bowtie$  bond $\bowtie$  atm] $\bowtie$  drug&(14 ; 9,317)&	14.20& 1 	&	0	&	0	&	1 (24 FDs)&0.0879&0.0030&0.0492\\
 \hline\hline
 
 \parbox[t]{2mm}{\multirow{4}{*}{\rotatebox[origin=c]{90}{{\bf PTC}}}}
&	atom $\bowtie$  molecule&(4 ; 9,111)&	13.67&	0.75	&	0.25	&	0	&	1 (4 FDs)&0.0231&0.0000&0.0015\\
&connected $\bowtie$  bond&(5 ; 24,758)&	1.50&	0.625	&	0.375	&	0	&	1 (8 FDs)&0.1012&0.0015&0.0015\\
&	[connected $\bowtie$  bond] $\bowtie$  molecule	&(6 ; 18,312) &27.08&	0.75	&	0.25	&	0	&	12 (12FDs) &0.0686&0.0000&0.0030\\
&	connected $\bowtie_{id1}$ [atom $\bowtie$  molecule]&(6 ; 18,312)&	27.08&	0.583	&	0.417	&	0	&1 (12 FDs)&0.0903&0.0030&0.0045\\
 \hline\hline

\parbox[t]{2mm}{\multirow{4}{*}{\rotatebox[origin=c]{90}{{\bf MIMIC3}}}} 
&	diagnosesicd $\bowtie$  patients&(12 ; 651,047)&	7.50&	0.591	&	0.273	&	0.136	& 1	(22 FDs) &2.1876&0.0015&2.3120\\
&	dicddiagnoses $\bowtie$  diagnosesicd&(7 ; 658,498)	&22.84&	0.333	&	0	&	0.667	&	1	(12 FDs)&1.7202&0.0000&0.3497\\
&	[diagnosesicd $\bowtie$ patients] $\bowtie$  dicddiagnoses&(14 ; 658,498)&	22.84&	0.545	&	0	&	0.455	&	1	(44 FDs)&5.1232&0.0000&6.1325\\
&Q(patients $\bowtie$  admissions)&(10 ; 6,736)&	0.79  &0.563 &0 &0437 &1	(16 FDs)&0.2360&0.0015&0.0230\\
 \hline\hline
 
 \parbox[t]{2mm}{\multirow{4}{*}{\rotatebox[origin=c]{90}{{\bf TPC-H}}}}
& Q2$^*$(P $\bowtie$ PS $\bowtie$ S $\bowtie$ N $\bowtie$ R)  &(10 ; 21,696) &	1.50& 0.594&0.087&0.319 &1	(69 FDs)&0.1299&0.0045&0.0120\\
& Q3$^*$(C $\bowtie$ O $\bowtie$ L) 
&(6 ; 60,150)&0.12&0.429 &0 &0.571&1	(14 FDs)&13.036&0.0000&0.0198\\
 
& Q9$^*$(P $\bowtie$ PS $\bowtie$ S $\bowtie$ L $\bowtie$ O $\bowtie$ N)   &(9 ; 3,735,632)&25,813& 0.875& 0.125 & 0&1	(8 FDs)&16.967&0.0015&12.1261\\
& Q11$^*$(P  $\bowtie$ S $\bowtie$ N)  &(15 ; 284,160)&80.09&0.636 &0.232 &0.132&1	(151 FDs)&13.771&0.0246&0.5777\\
 \hline
\end{tabular}
\caption{ \textRev{Accuracy and time breakdowns }of \jedi algorithms}
\label{exp:accuracyAndBreakdown}
\end{table*}

In Fig.~\ref{fig:jedi-time-repartition},  we report the average runtime breakdown of each algorithm \texttt{upstageFDs}, \texttt{inferFDs}, and \texttt{mineFDs} of \jedi (as horizontal histograms) and their respective percentages of discovered FDs (in the corresponding pie charts with the same color coding). Error bars represent the standard deviation of the average total runtime of \jedi algorithms. \texttt{selectionFDs}'s time is included in \texttt{upstageFDs}'s time.  Various base table data distributions, SPJ views, and coverage values with very different characteristics across the datasets illustrate the behavior of our algorithms.

Noticeably, \texttt{upstageFDs} can retrieve from 33.3\% to 59\% of the FDs in MIMIC3 in 0.75ms on average, from 58.3\% to 75\% in PTC in 1.12ms, from 66.7\% to 100\% in PTE in 1.5ms, and from 42.8.7\% to 87.5\% in TPC-H in 7.65ms, which are  negligible times compared to \texttt{mineFDs} time and I/O time, the most time-consuming steps. Table III gives the number of FDs retrieved by each algorithm and time breakdowns. 

 The logical inference times are negligible (below 0.0001 ms) and are not reported in the table.  \texttt{inferFDs} can   retrieve 100\% of the FDs for  3 queries over PTE. In some cases,  \texttt{mineFDs} is executed but does not return any new FD (e.g., Q$^*$9 in TPC-H or \texttt{[atm $\bowtie$ bond $\bowtie$ atm]$\bowtie$drug}), whereas in the other queries, mining subsets of the SPJ view using \texttt{mineFDs}  is necessary to recover the remaining FDs. 
 Without computing and mining the join results, \texttt{upstageFDs} and  \texttt{inferFDs} can retrieve 83.01\% of FDs  (68.38$\pm$20.00 and 14.63$\pm$15.47, respectively) on average across all the datasets.  \textRev{We also observed that join ordering does not affect the total number of discovered FDs, but it changes FD  provenance. The main reason is due to the difference in the query execution sub-trees triggering different FD validation by each \jedi algorithm.}

It should be noted that, due to the NP-completeness of the FD mining problem (cf. \cite{davies1994np}), this exploration is necessary to ensure that no valid FD is missed.

These results clearly show the main advantages of our approach, reducing drastically the execution time of FD discovery from SPJ queries with minimal memory consumption, outperforming all the state-of-the art methods tested in our experiments over a large representative range of queries.

\section{Related Work} 
\label{sec:relatedwork}
In the last three decades, numerous approaches from the database and the data mining communities have been proposed to extract automatically valid exact  and approximate FDs from single relational tables~\cite{KiMa95,caruccio2015relaxed}. Liu et al. \cite{Liu2012} have shown that the complexity of FD discovery is in $O(n^2(\frac{k}{2})^22^k)$ where $k$ is the number of attributes and $n$ the number of records considered.   %
 To find FDs efficiently, existing approaches can be classified into three categories:~(1) Tuple-oriented methods (e.g., FastFDs~\cite{WGR01}, DepMiner~\cite{lopes2000efficient}) that exploit the notion of tuples agreeing on the same values to determine the combinations of attributes of an FD; (2) Attribute-oriented methods (e.g., Tane~\cite{HKPT98, HKPT99}, Fun~\cite{NoCi01icdt, NoCi01is}, FDMine~\cite{YH08}) that  use pruning techniques and reduce the search space to the necessary set of attributes of the relation to discover exact and approximate FDs.    HyFD~\cite{HyFD} exploits simultaneously the tuple- and attribute-oriented approaches to outperform the previous approaches; and more recently 
(3) Structure learning methods relying on sparse regression \cite{Zhang2020}, or on entropy-based measures \cite{kenig2019mining} to score candidate constraints (not limited to FDs alone). More particularly, FDX \cite{Zhang2020}  performs structure learning over a sample constructed by taking the value differences over sampled pairs of tuples from the raw data. %
In addition, incremental approaches~(e.g., \cite{SchirmerP0NHMN19, CaruccioCDP19}) have been developed to tackle data volume and velocity with updating all valid FDs when new tuples are inserted outperforming classical approaches that recalculate all FDs after each data update. Extensive evaluations of FD discovery algorithms can be found in \cite{DurschSWFFSBHJP19,PapenbrockEMNRZ15}. To the best of our knowledge, previous work on FD discovery did not attempt to address the problem of FD discovery from integrated views in an efficient manner while preserving data provenance. Our approach combining logical inference  and selective mining from the base tables \textRev{adapting FD lattice approach} avoids the full computation of FDs from the integrated views and it is the first solution in this direction. 

\textRev{The problem of deciding whether a semantic constraint (being a FD or a Join Dependency) is valid on a tableau view, knowing that it is valid on the base relations has been addressed by Klug et al. \cite{KlugP82}. However, their view constraint problem and FD implication is inherently different from our FD inference problem, and allowing to reuse discovered FDs of base tables while annotating them with provenance information and efficiently recomputing FDs that are valid on the view and are not valid on the base tables (as stated in Th. 5). Moreover, the underlying technique employed in \cite{KlugP82} and used for checking the validity of FDs leverages the tableau chase as opposed to using the FD lattice, the latter being a well established efficient method for FD discovery.

The problem of propagating XML keys to relations is an orthogonal problem with respect to ours \cite{DavidsonFH07}. The simple mapping language from XML to relations and the restriction to XML keys, that cannot capture relational functional dependencies, is specific to this work as also stated in their paper \cite{DavidsonFH07}.}
\section{Conclusions}
\label{sec:conclusion}
We addressed the problem of FD discovery from integrated views starting from the FDs of multiple base tables by avoiding the full computation of the view beforehand. The salient features of our work are the following: (1) We leverage single-table approximate FDs that become exact FDs due to the join operation; (2) We leverage logical inference to discover FDs from the base tables  without computing the full view result; and (3) We find new multi-table join FDs from partial join using selective mining on the necessary attributes. %
 We empirically show that \jedi outperforms,  both in terms of runtime and memory consumption, the state-of-the-art FD discovery methods applied to the SPJ views that have to be computed beforehand.  
We hope that our work will open a new line of research for reusing the FDs discovered from multiple base tables. 
Various orderings of the base tables lead to different sets of potential upstaged FDs, which, in turn,  may trigger different logical inferences. 
 Future work will be to find the optimal ordering of the base tables to reduce the overall execution time and memory consumption. 

\bibliographystyle{abbrv}
\bibliography{biblio}

\clearpage

\appendix{Theorem and sketches of proofs:}

\setcounter{theorem}{0}
\setcounter{lemma}{0}
\begin{theorem}\label{thmApp:fdsConservation}\thmFont
    Let $\viewSpecArg{}$, $\viewSpecArg{1}$, and $\viewSpecArg{2}$ be view specifications.
    Let \fdSet, $\fdSet_1$ and $\fdSet_2$ be the sets of FDs over views specified by $\viewSpecArg{}$, $\viewSpecArg{1}$ and $\viewSpecArg{2}$, respectively.
    Let $\fds(V)$ denotes the set of FDs over a view specified by $V$.\\
    Then:
    \begin{align*}
        &\fds(\pi_X(\viewSpecArg{})) \subseteq \fdSet,  \qquad \fds(\sigma_\rho(\viewSpecArg{})) \supseteq \fdSet, \text{ and} \\[0.25cm]
        &\fds(\viewSpecArg{1} \diamond \viewSpecArg{2}) \supseteq \fdSet_1 \cup \fdSet_2 \text{, with }\diamond\in \{\bowtie;\fullouterjoin;\leftouterjoin;\rightouterjoin;\ltimes;\rtimes\}
    \end{align*}
\end{theorem}
\begin{IEEEproof}
Trivial.
\end{IEEEproof}

\begin{lemma}[FD types conservation though join operations]
Let $\relationSet = \{R_1;\dots;R_n\}$ be a set of relational instances.
Let $\fds(R_i)$ denotes the set of minimal FDs over the relationnal instance $R_i$.
Let $\Sigma_{left\_up},\Sigma_{right\_up}, \Sigma_{inf} \text{and} \Sigma_{join}$ denote the sets of left upstaged, right upstaged, inferred and join FDs over $R_1 \diamond\dots\diamond R_n$, respectively.
Then, the sets $\Sigma_{left\_up}\cup\Sigma_{right\_up}, \Sigma_{inf} \text{ and } \Sigma_{join}$ are equals regardless of the join order.
At the opposite, equality between sets $\Sigma_{left\_up}$ and $\Sigma_{right\_up}$ cannot be guaranteed.

\end{lemma}
\begin{IEEEproof}

\end{IEEEproof}

\begin{lemma}[Upstaged join FDs]\thmFont
 Let $\leftInst$ and $\rightInst$ be two instances over relations $\leftSch$ and $\rightSch$, respectively, and $\fdSet_\leftInst$ and $\fdSet_\rightInst$ be the two sets of all FDs such that $\leftInst \models \fdSet_\leftInst$ and $\rightInst \models \fdSet_\rightInst$, respectively. \\ Then the sets of upstaged FDs denoted $\fdSet^{new}_{{\leftInst}}$ and $\fdSet^{new}_{{\rightInst}}$ are the sets:
  \begin{align}
      \fdSet^{new}_{{\leftInst}} &= \{ d\ |\ d\not\in\fdSet_{\leftInst} \wedge \left(\leftInst \Diamond_{\leftJoinAtt=\rightJoinAtt} (\pi_{\rightJoinAtt}(\rightInst))\models d \right)\} \\
    \fdSet^{new}_{{\rightInst}} &= \{ d\ |\ d\not\in\fdSet_{\rightInst} \wedge \left(\pi_{\leftJoinAtt}(\leftInst) \Diamond_{\leftJoinAtt=\rightJoinAtt} \rightInst)\models d \right)\}
  \end{align}
\end{lemma}

\begin{lemma}\label{lemmaApp:notModelsImplication}\thmFont
Let $\leftInst$ and $\rightInst$ be two instances over relations $\leftSch$ and $\rightSch$, respectively. Let $\leftInst \Diamond_{\leftJoinAtt=\rightJoinAtt} \rightInst$ be a join result with $\leftJoinAtt \subseteq \atts(\leftInst)$, $\rightJoinAtt  \subseteq \atts(\rightInst)$. For all $A  \subseteq \atts(\leftInst)\setminus \leftJoinAtt$ and $B \subseteq \atts(\rightInst)\setminus \rightJoinAtt$:
$$\text{if }\leftInst \Diamond_{\leftJoinAtt=\rightJoinAtt} \rightInst \not\models \leftJoinAtt \to B \text{ then } \leftInst \Diamond_{\leftJoinAtt=\rightJoinAtt} \rightInst \not\models A \to B $$
\end{lemma}

\begin{IEEEproof}[Proof sketch]
Let $x_1,\dots,x_n$ being values over the attributes in $\leftJoinAtt$, and $b_1,\dots,b_m,b^\prime_1,\dots,b^\prime_m$ being values over the attributes in $B$. If $\leftInst \Diamond_{\leftJoinAtt=\rightJoinAtt} \rightInst \not\models \leftJoinAtt \to B $ then there exist two tuples:
\begin{align*}
    &t_\rightInst(x_1,\dots,x_n,b_1,\dots,b_m,\dots)\\
    &t_\rightInst^\prime(x_1,\dots,x_n,b^\prime_1,\dots,b^\prime_m,\dots)
\end{align*}

in $\rightInst$ such that:
    $\exists i \in [1,\dots,m], b_i \neq b^\prime_i$ and
  there exists a tuple $t_\leftInst(x_1,\dots,x_n,a_1,\dots,a_k,\dots)$ in $\leftInst$ with $a_1,\dots,a_k$, the values of the attributes in $A$ (otherwise, tuples $t_\rightInst$ and $t_\rightInst^\prime$ would have been filtered during the join operation).
Thus, the join $\leftInst \Diamond_{\leftJoinAtt=\rightJoinAtt} \rightInst$  leads to the two tuples:
\[
\begin{split}
    t(x_1,\dots,x_n,a_1,\dots,a_k,b_1,\dots,b_m,\dots)\\ t^\prime(x_1,\dots,x_n,a_1,\dots,a_k,b^\prime_1,\dots,b^\prime_m,\dots)
\end{split}
\]
which violate the FD $A \to B$ and 
$\leftInst \Diamond_{\leftJoinAtt=\rightJoinAtt} \rightInst \not\models A \to B $
\end{IEEEproof}

\begin{theorem}\label{thmApp:MonoInstanceCrossJoinFDs}\thmFont
  Let $\leftInst$ and $\rightInst$ be two instances over relations $\leftSch$ and $\rightSch$, respectively. 
Let $\leftInst \Diamond_{\leftJoinAtt=\rightJoinAtt} \rightInst$ be a join result with $\leftJoinAtt \subseteq \atts(\leftInst)$, $\rightJoinAtt  \subseteq \atts(\rightInst)$. 
For all $A  \subseteq \atts(\leftInst)\setminus \leftJoinAtt$ and $B \subseteq \atts(\rightInst)\setminus \rightJoinAtt$,\\
 If $\leftInst \Diamond_{\leftJoinAtt=\rightJoinAtt} \rightInst \models A \to \leftJoinAtt \wedge  \leftInst \Diamond_{\leftJoinAtt=\rightJoinAtt} \rightInst \models \leftJoinAtt \to B$,\\
 Then $\leftInst \Diamond_{\leftJoinAtt=\rightJoinAtt} \rightInst \models A \to B$.
\end{theorem}
\begin{IEEEproof}
This is trivially derived from lemma~\ref{lemmaApp:notModelsImplication} by transitivity, with the use of Armstrong's transitivity axiom.
\end{IEEEproof}

\begin{theorem}\thmFont
Let $\leftInst$ and $\rightInst$ be two instances over relations $\leftSch$ and $\rightSch$, respectively. Let $\leftInst \Diamond_{\leftJoinAtt=\rightJoinAtt} \rightInst$ be a join result with $\leftJoinAtt \subseteq \atts(\leftInst)$, $\rightJoinAtt  \subseteq \atts(\rightInst)$. 
We cannot guarantee that all FDs over $\leftInst \Diamond_{\leftJoinAtt=\rightJoinAtt} \rightInst$ can be inferred from Armstrong's axioms over the FDs over $\leftInst$ and $\rightInst$ taken separately.
\end{theorem}
\begin{IEEEproof}
In the two following instances \leftInst and \rightInst, we can see that only the FDs $\rightJoinAtt A^\prime \to b$ and $\rightJoinAtt b \to A^\prime$ hold.
\begin{center}
    \begin{minipage}{0.25\columnwidth}
    \centering
        \begin{tabular}{cc}
            \multicolumn{2}{c}{$\leftInst$}\\
            \hline
            $\leftJoinAtt$ & $A$ \\
            \hline
            0 & 0\\
            1 & 0\\
            1 & 1\\
            2 & 2\\
            & \\
            & \\
        \end{tabular}
    \end{minipage}
    \begin{minipage}{0.25\columnwidth}
    \centering
        \begin{tabular}{ccc}
            \multicolumn{3}{c}{$\rightInst$}\\
            \hline
            $\rightJoinAtt$ & $A^\prime$ & $b$ \\
            \hline
            0 & 0 & 0\\
            1 & 0 & 0\\
            1 & 1 & 1\\
            2 & 1 & 0 \\
               & \\
            & \\
        \end{tabular}
    \end{minipage}
    \begin{minipage}{0.4\columnwidth}
\centering
\begin{tabular}{cccc}
    \multicolumn{4}{c}{${\leftInst \Diamond_{\leftJoinAtt=\rightJoinAtt} \rightInst}$}\\
    \hline
    $\leftJoinAtt=\rightJoinAtt$ & $A$ & $A^\prime$ & $b$\\
    \hline
    0 & 0 & 0 & 0\\
    1 & 0 & 0 & 0\\
    1 & 0 & 1 & 1\\
    1 & 1 & 0 & 0\\
    1 & 1 & 1 & 1\\
    2 & 2 & 1 & 0\\
\end{tabular}
\end{minipage}
\end{center}
In the join result, the FD $AA^\prime \to b$ holds but it cannot be inferred using Armstrong's axioms over the FDs discovered from each instance $\leftInst$ and $\rightInst$.
\end{IEEEproof}

\begin{theorem}\label{theo6App}\thmFont
 Let $\leftInst$ and $\rightInst$ be two instances over relations $\leftSch$ and $\rightSch$, respectively. 
Let $\leftInst \Diamond_{\leftJoinAtt=\rightJoinAtt} \rightInst$ be a join result with $\leftJoinAtt \subseteq \atts(\leftInst)$, $\rightJoinAtt  \subseteq \atts(\rightInst)$. 
For all $A  \subseteq \atts(\leftInst)$, $A^\prime  \subseteq \atts(\rightInst)$ and $b \in \atts(\rightInst)$:
If $\leftInst \Diamond_{\leftJoinAtt=\rightJoinAtt} \rightInst \models AA^\prime \to b$, Then $\leftInst \Diamond_{\leftJoinAtt=\rightJoinAtt} \rightInst \models \rightJoinAtt A^\prime \to b$.
\end{theorem}

\begin{theorem}[\jedi completeness]\thmFont
Let $\relationSet = \{R_1;\dots;R_n\}$ be a set of relational instances.
Let $\viewSpec$ be a view specification over relations in $\mathbf{R}$.
Let $\fds(\viewSpec)$ denotes the set of minimal FDs over the view specified by $\viewSpec$.
Let $\fds(\jedi)$ denotes the set of FDs computed by \jedi over the view specified by $\viewSpec$. Then: $$\forall d \in \fds(\viewSpec), \exists d^\prime \in \fds(\jedi) \text{ s.t. } d \equiv d^\prime$$
\end{theorem}
\begin{IEEEproof}[Proof sketch]
 Algorithm~\ref{alg:mainFramework} begins by extracting the attributes that are retrieved by a view in order to guide the mining process of these attributes while still guaranteeing completeness. %
Then Algorithm~\ref{alg:mainFramework} recursively explores the view specification, thus its completeness can be shown inductively.
For the base case where we have a relational instance $R$ (Algorithm~\ref{alg:mainFramework}, line\#~\ref{line:baseCase}), the FDs are mined using a classic level-wise algorithm allowing to retrieve every FD in $R$. In the case of a projection (Algorithm~\ref{alg:mainFramework}, line\#~\ref{line:projectionCase}), as the relevant attributes are  extracted at the very first step of Algorithm~\ref{alg:mainFramework}, no computation is performed and the previously computed FDs are returned. In the cases of a selection or a join (Algorithm~\ref{alg:mainFramework}, line\#~\ref{line:selectionCase} or line\#~\ref{line:joinCase} respectively), the tuples are filtered by the join operation and Algorithm~\ref{alg:upstagedSelFDs} mines all  exact FDs.  
Algorithm~\ref{algo:MonoInst_CrossJoinFDs} retrieves the minimal FDs from the logically inferred FDs, thus no \texttt{lhs} subset of FDs remains unchecked. Theorem~\ref{theo6} indicates which part of the candidate FD lattice is pruned. 
Then, Algorithm~\ref{algo:MultiInst_CrossJoinFDs} explores the candidate FDs using a classic bottom-up approach, thus no minimal FD remains unchecked. Overall, our algorithms explore the lattice of candidate FDs until they find minimal FDs; they avoid only the parts of the lattice that do not contain valid candidate FDs, thus \jedi with selective mining retrieves the complete set of minimal candidate FDs.
\end{IEEEproof}

\begin{theorem}[\jedi correctness]\thmFont
Let $\relationSet = \{R_1;\dots;R_n\}$ be a set of relational instances.
Let $\viewSpec$ be a view specification over relations in $\mathbf{R}$.
Let $I_\viewSpec$ be the view specified by $\viewSpec$.
Let $\fds(\jedi)$ denotes the set of FDs computed by \jedi over the view specified by $\viewSpec$. Then: $$\forall d \in \fds(\jedi), I_\viewSpec \models d$$
\end{theorem}
\begin{IEEEproof}[Proof sketch]
Algorithm~\ref{alg:mainFramework} recursively explores the view specification, thus its correctness can be shown inductively.
For the base case where we have a relational instance $R$ (Algorithm~\ref{alg:mainFramework}, line\#~\ref{line:baseCase}) the FDs are mined using a classic level-wise algorithm allowing to retrieve FDs in $R$. %

In case of a projection (Algorithm~\ref{alg:mainFramework}, line\#~\ref{line:projectionCase}), the process ensure that only the FDs with relevant attributes (i.e., the one that are retrieved in the output view) are kept. Thus, if the FDs output by the steps are correct, the handling of projection will only lead to remove irrelevant FDs without introducing incorrect ones.

In case of a selection (Algorithm~\ref{alg:mainFramework}, line\#~\ref{line:selectionCase}), Algorithm~\ref{alg:upstagedSelFDs}, valid exact FDs are discovered from the data using the principle of the level-wise algorithms adapted to mine upstaged FDs, thus the FDs are guaranteed to hold on the joined instance and only minimal FDs are kept.

In case of a join (Algorithm~\ref{alg:mainFramework}, line~\ref{line:joinCase}),
at first in Algorithm~\ref{alg:filteringFDs}, valid exact FDs are discovered from the data using the same principle as in Algorithm~\ref{alg:upstagedSelFDs}, thus returning FDs that are guaranteed to hold on the joined instance. Theorem~\ref{thm:MonoInstanceCrossJoinFDs} shows the correctness of the set of FDs inferred through logical inference. Then, the subroutine \subRefine checks the correctness of its candidates FDs holding on the data. Therefore, the set of FDs $\fdSet_{2}$ discovered by Algorithm~\ref{algo:MonoInst_CrossJoinFDs} is such that  $\fdSet_{2} \models \fdSet_{\Diamond}$.  In Algorithm~\ref{algo:MultiInst_CrossJoinFDs}, Theorem~\ref{theo6} enforces the retrieval of FDs based only on the attributes that can become \texttt{rhs} in the joined instance, then only plausible candidate FDs are explored. Therefore, every discovered FD holds in the joined instance. By construction, Algorithm~\ref{algo:MonoInst_CrossJoinFDs} and \ref{algo:MultiInst_CrossJoinFDs} lead to the retrieval of FDs that are minimal.
\end{IEEEproof}
\end{document}